\documentclass[aps,prd,showpacs,notitlepage,nofootinbib,floatfix,showkeys,superscriptaddress]{revtex4-2}
\usepackage{blindtext}
\usepackage{hyperref}
\usepackage{amsmath,amssymb}
\usepackage{float}
\usepackage{microtype}
\usepackage{graphicx}
\usepackage{bm}
\usepackage{latexsym}
\usepackage{epsfig}
\usepackage{psfrag}
\usepackage{color}
\usepackage[dvipsnames]{xcolor}
\usepackage{subfigure}
\usepackage{graphicx}

\usepackage{amssymb}
\usepackage{amsmath}
\usepackage{bm}
\usepackage{latexsym}
\usepackage{epsfig}
\usepackage{psfrag}
\usepackage[normalem]{ulem}
\usepackage{textcomp}
\usepackage{color}
\usepackage{pstricks}
\usepackage[utf8]{inputenc}

\def\nn{\nonumber} 
\def\f{\frac}
\def\l{\left}
\def\r{\right}
\def\d{{\mathrm{d}}}
\def\pa{{\partial}}
\def\Mpl{M_{_{\mathrm{Pl}}}}
\def\Mp{M_{_{\mathrm{Pl}}}}
\def\vk{{\bm k}}

\def\vka{{\bm k}_{1}}
\def\vkb{{\bm k}_{2}}
\def\vkc{{\bm k}_{3}}
\def\ska{{k_{1}}}
\def\skb{{k_{2}}}
\def\skc{{k_{3}}}

\def\cG{{\mathcal{G}}}
\def\cR{{\mathcal{R}}}
\def\ei{\eta_{\mathrm{i}}}

\def\ee{\eta_{\mathrm{e}}}
\def\fnl{f_{_{\mathrm{NL}}}}

\def\ki{k_{\mathrm{i}}}

\def\mpcinv{{\mathrm{Mpc}^{-1}}}
\def\ps{{\mathcal{P}}_{_{\mathrm{S}}}}
\def\pt{{\mathcal{P}}_{_{\mathrm{T}}}}
\def\As{A_{_{\mathrm{S}}}}
\def\ns{n_{_{\mathrm{S}}}}

\allowdisplaybreaks

\begin{document}
%%%%%%%%%%%%%%%%%%%%%%%%%%%%%%%%%%%%%%%%%%%%%%%%%%%%%%%%%%%%%%%%%%%%%%%%%%%%%%
\title{Suppression of scalar power on large scales and associated bispectra}
\author{H.~V.~Ragavendra}
\email{Current address: Department of Physical Sciences, Indian Institute of 
Science Education and Research Kolkata, Mohanpur, Nadia 741246, India.
E-mail: ragavendra.pdf@iiserkol.ac.in} 
\affiliation{Centre for Strings, Gravitation and Cosmology, Department of Physics, 
Indian Institute of Technology Madras, Chennai~600036, India}
\author{Debika Chowdhury}
\email{E-mail: debika.chowdhury@swansea.ac.uk}
\affiliation{Department of Theoretical Physics, Tata Institute of 
Fundamental Research, Mumbai~400005, India}
\affiliation{Department of Physics, Swansea University, Swansea, SA2 8PP, U.K.}
\author{L.~Sriramkumar}
\email{E-mail: sriram@physics.iitm.ac.in}
\affiliation{Centre for Strings, Gravitation and Cosmology, 
Department of Physics, Indian Institute of Technology Madras, Chennai~600036, India}
%%%%%%%%%%%%%%%%%%%%%%%%%%%%%%%%%%%%%%%%%%%%%%%%%%%%%%%%%%%%%%%%%%%%%%%%%%%%%%%
\begin{abstract}
A sharp cut-off in the primordial scalar power spectrum on large scales 
has been known to improve the fit to the cosmic microwave background (CMB) 
data when compared to the more standard, nearly scale invariant power spectra
that arise in slow roll inflation.
Over the last couple of years, there has been a resurgent interest in arriving 
at such power spectra in models with kinetically dominated initial conditions
for the background scalar field which leads to inflation of specific duration.
In an earlier work, we had numerically investigated the characteristics of the 
scalar bispectrum generated in such models.
In this work, we compare the scenario with two other competing scenarios 
(viz. punctuated inflation and a model due to Starobinsky) which also 
suppress the scalar power in a roughly similar fashion on large scales.
We further consider two other scenarios involving inflation 
of a finite duration, one wherein the scalar field begins on the inflationary 
attractor and another wherein the field starts with a smaller velocity and 
evolves towards the attractor.
These scenarios too exhibit a sharp drop in power on large scales if the
initial conditions on the perturbations for a range of modes are imposed 
on super-Hubble scales as in the kinetically dominated model.
We compare the performance of all the models against the Planck CMB data 
at the level of scalar power spectra.
The model wherein the background field always remains on the inflationary 
attractor is interesting for the reason that it permits analytical calculations 
of the scalar power and bi-spectrum.
We also compare the amplitudes and shapes of the scalar non-Gaussianity 
parameter~$\fnl$ in all these cases which lead to scalar power spectra 
of similar form.
Interestingly, we find that, in the models wherein the initial conditions
on the perturbations are imposed on super-Hubble scales, the consistency 
relation governing the scalar bispectrum is violated for the large scale 
modes, whereas the relation is satisfied for all the modes in the other 
scenarios.
These differences in the behavior of the scalar bispectra can conceivably 
help us observationally discriminate between the various models which 
lead to scalar power spectra of roughly similar shape.
\end{abstract}
\maketitle

%%%%%%%%%%%%%%%%%%%%%%%%%%%%%%%%%%%%%%%%%%%%%%%%%%%%%%%%%%%%%%%%%%%%%%%%%%%%%%%%

\section{Introduction}

Ever since the advent of the three-year WMAP data, it has been repeatedly
found that a sharp drop in power at large scales roughly corresponding to 
the Hubble radius today improves the fit to the anisotropies in the cosmic 
microwave background (CMB) at the low multipoles (for an early analysis,
see, for instance, Ref.~\cite{Bridle:2003sa}; for later discussions in this 
context, see Refs.~\cite{Shafieloo:2003gf,Hunt:2004vt,Hunt:2007dn,
Hazra:2013nca}).
A variety of inflationary scenarios have been constructed to generate such 
a drop in power on large scales (for a short list of possibilities, 
see Refs.~\cite{Cline:2003ve,Contaldi:2003zv,Sinha:2005mn,Powell:2006yg,
Boyanovsky:2006pm,Boyanovsky:2006qi,Nicholson:2007by,Jain:2007au,Jain:2008dw,
Jain:2009pm,Hazra:2014jka,Hazra:2014goa,White:2014aua,Qureshi:2016pjy,Pi:2017gih}).

One of the scenarios that generates a scalar spectrum with suppressed power 
on large scales corresponds to a situation wherein the scalar field driving 
inflation starts rolling down the potential with a high velocity (for the 
original discussion, see Ref.~\cite{Contaldi:2003zv}; for more recent 
discussions, see Refs.~\cite{Ramirez:2011kk,Ramirez:2012gt,Cicoli:2014bja,
Hergt:2018crm,Hergt:2018ksk}).
While the very early kinetically dominated phase does not permit accelerated
expansion, the friction arising due to the expansion of the universe slows 
down the field, initially leading to a brief period of fast roll inflation
and eventually to the standard phase of slow roll inflation.
If one chooses the beginning of inflation to occur at an appropriately early 
time, the inflationary power spectra exhibit lower power at suitably large 
scales, improving the fit to the CMB data at the low multipoles~\cite{Hergt:2018ksk}.
However, it should be emphasized that, in such scenarios, a range of large scale
modes are never inside the Hubble radius and spectra with a suppression of 
power arise provided the standard Bunch-Davies initial conditions are 
imposed on super-Hubble scales~\cite{Contaldi:2003zv,Hergt:2018ksk}.

A competing inflationary scenario that, in fact, leads to sharper drop 
in power at the large scales corresponds to a short phase of fast roll 
sandwiched between two epochs of slow roll inflation.
Such scenarios can be further sub-divided into two categories:~one wherein 
inflation is sustained even during the phase of fast roll and another 
wherein the epoch of fast roll leads to a brief departure from inflation.
While the first type of scenario can be achieved in a model originally 
due to Starobinsky involving a linear potential with an abrupt change 
in slope~\cite{Starobinsky:1992ts}, the second type of scenario---dubbed 
punctuated inflation---is known to arise due to inflationary potentials 
containing a point of inflection~\cite{Jain:2008dw,Jain:2009pm}.
The advantage of such scenarios is that the initial epoch of slow roll 
inflation permits one to impose the standard Bunch-Davies initial 
conditions in the sub-Hubble domain for {\it all}\/ the modes of 
cosmological interest.

As we shall see, these alternative scenarios lead to scalar power spectra 
which have almost the same shape.
Moreover, as we shall illustrate, these power spectra also lead to a 
slightly improved fit to the CMB data than the nearly scale invariant 
power spectra.
One can expect that non-Gaussianities, specifically, the scalar bispectrum, 
would help us discriminate between these models.
In an earlier work, we had numerically computed the scalar bispectrum and 
the corresponding non-Gaussianity parameter~$\fnl$ that arise in models 
with kinetically dominated initial conditions~\cite{Ragavendra:2019mek}.
Interestingly, we had found that, in such a scenario, the contributions 
due to the boundary terms in the third order action governing the scalar
perturbations dominate the contributions due to the bulk terms.
In this work, we shall discuss in detail the various contributions to the 
scalar bispectrum that arise as well as the numerical procedure that we have
adopted to compute the scalar bispectrum.
We shall also compare the bispectrum that arises in the model with those 
that occur in the Starobinsky model and punctuated inflation. 
Moreover, apart from the above mentioned scenarios, we shall also examine 
two other situations involving inflation of a finite duration, which can 
be considered to be variations of the model with kinetically dominated 
initial conditions.
We shall consider a case wherein the background scalar field begins
on the inflationary attractor (a scenario which we shall call as the 
hard cut-off model) and another wherein the field starts with a small 
velocity and evolves towards the attractor (a scenario which we shall 
refer to as the dual to kinetic domination).
As in the model with an early kinetically dominated phase, these cases too 
lead to a sharp drop in power on large scales when the initial conditions 
on the perturbations are imposed on super-Hubble scales for a range of
modes. 
Further, since the trajectory always remains on the attractor in the 
hard cut-off model, it leads to slow roll, permitting us to evaluate the 
scalar power and bispectra analytically.
We find that, in the equilateral limit, the amplitude of the scalar 
non-Gaussianity parameter $\fnl$ proves to be very large [with $\fnl 
\simeq \mathcal{O}(10^2\mbox{--}10^6)$] in the scenario of dual to 
kinetic domination and the hard cut-off model.
We also find that $\fnl$ in the equilateral limit is relatively larger in 
the model with kinetically dominated initial conditions 
[with $\fnl \simeq \mathcal{O}(1\mbox{--}10)$] as well as in the
Starobinsky model (with $\fnl \simeq 10$).
Moreover, as expected, in the models wherein the Bunch-Davies initial 
conditions are imposed on super-Hubble scales, the consistency relation 
governing the scalar bispectrum is violated for the large scale modes, 
whereas the relation is satisfied for all the modes in the other 
scenarios (viz. the Starobinsky model and punctuated inflation).
These differences in the behavior of the scalar bispectrum can hopefully 
help us observationally discriminate between the various models.

The remainder of this paper is organized as follows. 
In the next section, we shall discuss the power spectra that arise in the 
different inflationary scenarios of interest, viz. inflation with kinetically 
dominated initial conditions, the Starobinsky model, punctuated inflation,
the hard cut-off model and the model which is dual to kinetic domination.
We shall also compare the scalar and tensor power spectra with the CMB data.
In Sec.~\ref{sec:toa}, we shall discuss the third order action governing the 
curvature perturbation, including the boundary terms.
In Sec.~\ref{sec:sbs}, we shall numerically evaluate the scalar bispectra that
arise in all these models.
We shall also present the analytical calculation of the scalar bispectrum in
the hard cut-off model.
In Sec.~\ref{sec:fnl}, we shall describe the amplitude and the shape of the  
scalar non-Gaussianity parameter $\fnl$ that arise in all the cases.
In Sec.~\ref{sec:cr}, we shall examine the consistency relation governing the 
scalar bispectrum in the squeezed limit.
We shall conclude in Sec.~\ref{sec:dis} with a summary of the results obtained.
In two appendices, we shall illustrate the imprints of the initial kinetically 
dominated epoch on the scalar power spectrum across different inflationary 
models and discuss the constraints on the cosmological parameters in some
specific models.

A few words on our conventions and notations are in order at this stage of our
discussion.
We shall work with natural units wherein $\hbar=c=1$, and define the Planck 
mass to be $\Mpl=(8\,\pi\, G)^{-1/2}$. 
We shall adopt the signature of the metric to be $(-,+,+,+)$ and assume the
background to be the spatially flat Friedmann-Lema\^itre-Robertson-Walker~(FLRW) 
line element described by the scale factor $a$ and the Hubble parameter~$H$. 
An overdot and an overprime shall represent differentiation with respect 
to the cosmic time ($t$) and the conformal time ($\eta$) coordinates,
respectively.
Further, we shall denote the number of $e$-folds by~$N$.

%%%%%%%%%%%%%%%%%%%%%%%%%%%%%%%%%%%%%%%%%%%%%%%%%%%%%%%%%%%%%%%%%%%%%%%%%%%%%%%

\section{Suppressing the scalar power on large scales}\label{sec:lpq}

In this section, we shall describe the models of our interest and discuss 
the scalar power spectra that arise in these cases.

%%%%%%%%%%%%%%%%%%%%%%%%%%%%%%%%%%%%%%%%%%%%%%%%%%%%%%%%%%%%%%%%%%%%%%%%%%%%%%%

\subsection{Numerical evaluation of the scalar power spectrum}

Let us begin by describing the evaluation of the scalar power spectrum in 
inflation driven by a single, canonical, scalar field.
Recall that, in such a case, the evolution of the scalar perturbations 
is governed by the following equation of motion for the Mukhanov-Sasaki 
variable $v_k$ (see, for instance, the reviews~\cite{Mukhanov:1990me,
Martin:2003bt,Martin:2004um,Bassett:2005xm,Sriramkumar:2009kg,
Sriramkumar:2012mik,Baumann:2009ds,Linde:2014nna,Martin:2015dha}):
\begin{equation}
v_k'' + \l(k^2 - \f{z''}{z}\r)\,v_k = 0,\label{eq:mse}
\end{equation}
where $z=\sqrt{2\,\epsilon_1}\,\Mp\,a$, with $\epsilon_1$ being the first 
slow roll parameter defined as $\epsilon_1=-\dot{H}/H^2$.
The scalar power spectrum $\ps(k)$ is given by
\begin{equation}
\ps(k) =\f{k^3}{2\,\pi^2}\, \vert f_k\vert^2
=\f{k^3}{2\,\pi^2}\,\l(\f{\vert v_k\vert}{z}\r)^2,\label{eq:ps}
\end{equation}
where we have introduced the quantity 
\begin{equation}
f_k=\f{v_k}{z},\label{eq:fk-vk}
\end{equation} 
which denotes the Fourier modes associated with the curvature perturbation.
Usually the Bunch-Davies initial conditions are imposed on the variable $v_k$
at early times in a domain wherein $k \gg \sqrt{z''/z}$.
The modes are evolved from these initial conditions, and the power spectra are 
evaluated at late times such that $k \ll \sqrt{z''/z}$.
In the conventional slow roll inflationary scenario, these conditions correspond
to the modes being in the sub-Hubble [i.e. when $k\gg (a\,H)$] and the 
super-Hubble [i.e. when $k\ll (a\,H)$] domains, respectively.
While, analytically, one imposes the Bunch-Davies conditions in the limit $k\gg 
(a\,H)$, numerically, one often finds that it is adequate if the initial
conditions on the perturbations are imposed when $k\simeq 10^2\, (a\, H)$.
Moreover, theoretically, the spectra are to be evaluated in the super-Hubble
limit~$k\ll (a\,H)$.
However, other than in a few peculiar models, the amplitude of the curvature 
perturbation $f_k$ quickly freezes once the modes leave the Hubble radius. 
Due to this reason, the power spectra are numerically evaluated typically 
when $k\simeq 10^{-5}\, (a\, H)$ (see, for instance, 
Refs.~\cite{Chen:2008wn,Hazra:2012yn}).

When comparing with the CMB data, we shall also evaluate the tensor 
power spectra and take into account their contributions.
Note that the Mukhanov-Sasaki variable associated with the tensor
perturbations satisfy the differential equation~\cite{Mukhanov:1990me,
Martin:2003bt,Martin:2004um,Bassett:2005xm,Sriramkumar:2009kg,
Sriramkumar:2012mik,Baumann:2009ds,Linde:2014nna,Martin:2015dha}
\begin{equation}
u_k'' + \l(k^2 - \f{a''}{a}\r)\,u_k = 0,\label{eq:mse-t}
\end{equation}
and the tensor power spectrum $\pt(k)$ is defined as 
\begin{equation}
\pt(k)= \f{8}{\Mpl^2}\, \f{k^3}{2\,\pi^2}\,\l(\f{\vert u_k\vert}{a}\r)^2.
\end{equation}
While evaluating the tensor power spectrum numerically, as in the scalar 
case, the standard Bunch-Davies initial conditions can be imposed on the 
perturbations when $k\simeq 10^2\, (a\, H)$ and the spectrum can be 
evaluated when $k\simeq 10^{-5}\, (a\, H)$.
%%%%%%%%%%%%%%%%%%%%%%%%%%%%%%%%%%%%%%%%%%%%%%%%%%%%%%%%%%%%%%%%%%%%%%%%%%%%%%%

\subsection{The models of interest}

Let us now describe the different inflationary models that we shall consider 
and discuss the scalar power spectra arising in these models.

%%%%%%%%%%%%%%%%%%%%%%%%%%%%%%%%%%%%%%%%%%%%%%%%%%%%%%%%%%%%%%%%%%%%%%%%%%%%%%%

\subsubsection{Models with kinetically dominated initial conditions}

The scenario of our primary interest is the one with kinetically 
dominated initial conditions, i.e. the situation wherein the kinetic 
energy of the inflaton completely dominates its potential energy 
during the initial stages of evolution~\cite{Contaldi:2003zv,
Horner:2013tma,Handley:2014bqa,Horner:2014bba,Hergt:2018crm,
Hergt:2018ksk}.
We shall examine the scenario in the quadratic potential (which
we shall refer to as QP)
\begin{equation}
V(\phi)=\f{1}{2}\,m^2\,\phi^2,\label{eq:qp}  
\end{equation}
and the Starobinsky model described by the potential 
\begin{equation}
V(\phi)=\f{\Lambda}{8}\,
\l[1-{\rm exp}\l(-\sqrt{\f{2}{3}}\,\f{\phi}{\Mpl}\r)\r]^2.
\label{eq:sm1}
\end{equation}
As we shall also be considering a different model due to
Starobinsky, we shall refer to the model described by the
above potential as Starobinsky model~I (or, simply, SMI,
hereafter). 

In the above potentials, to achieve kinetic domination, we shall 
set the initial value of the first slow roll parameter to be 
$\epsilon_{1{\mathrm{i}}}=2.99$.
Evidently, this value determines the initial velocity of the field.
The expansion of the universe slows down the field and one finds that 
inflation sets in (i.e. $\epsilon_1$ becomes less than unity) after 
about an $e$-fold or two (say, at $N_1$) when counted from, say, $N=0$, 
when we begin evolving the background.
Moreover, slow roll inflation (say, when $\epsilon_1\lesssim 10^{-2}$) 
is actually achieved only after a few $e$-folds.
We shall choose the initial value of the field so as to lead to 
adequate number of $e$-folds (say, about~$60$ or so) before inflation
is terminated at late times.
We shall assume that the pivot scale of $k_\ast\simeq 
5\times10^{-2}\, \mathrm{Mpc}^{-1}$ leaves the Hubble radius at
$N_\ast$ number of $e$-folds {\it prior to the end of inflation}.\/
As we shall discuss later, we shall be comparing the scalar power
spectra from the different inflationary models with the CMB data.
When doing so, we shall vary $N_\ast$, along with the inflationary 
parameters, to arrive at the best-fit values for $N_\ast$ as well 
as the other parameters.

\begin{figure}[!t]
\begin{center}
\includegraphics[width=15.0cm]{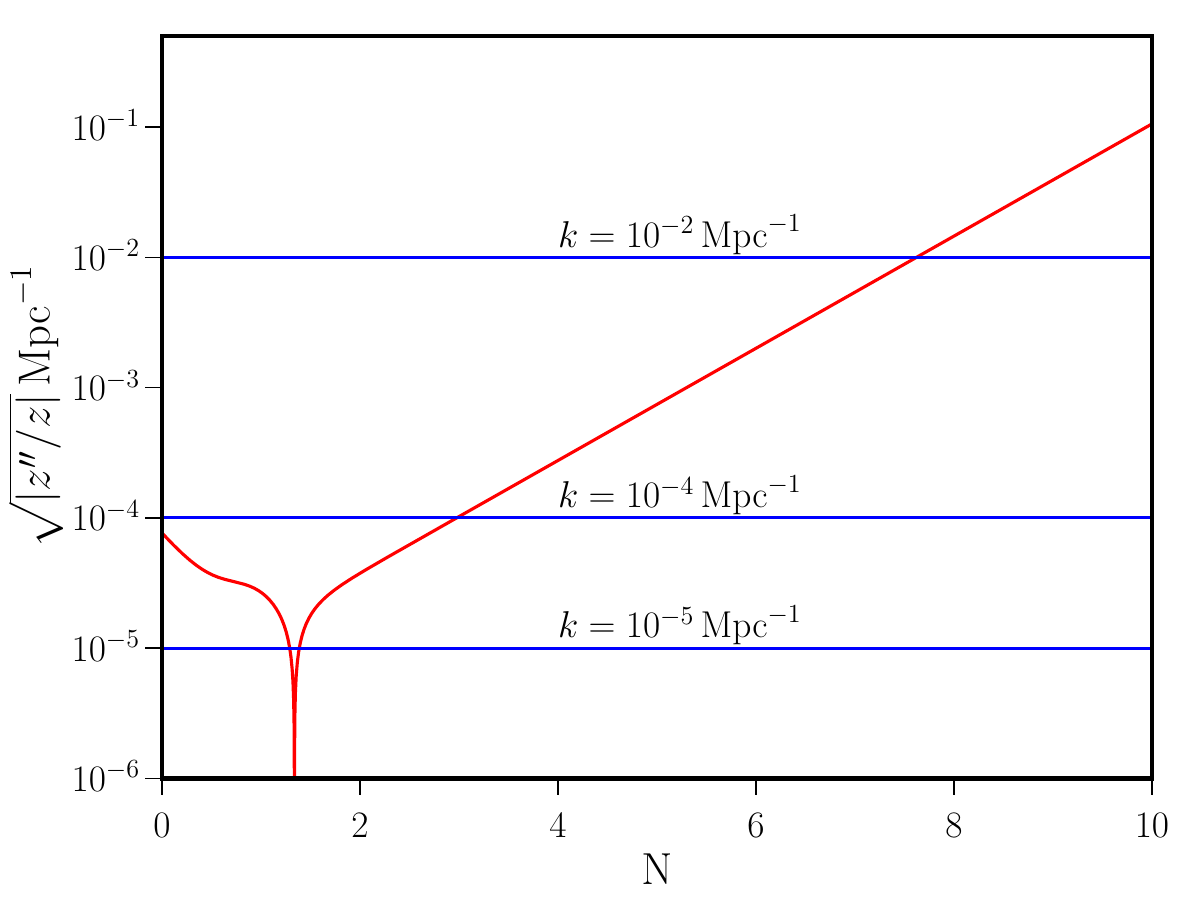}
\end{center}
\vskip -15pt
\caption{The behavior of the quantity $\sqrt{\vert z''/z\vert}$ 
has been plotted (in red) as a function of $e$-folds $N$ in an inflationary 
scenario of finite duration achieved due to an initial epoch of kinetic 
domination in the model which we refer to as QPa.
Note that $\sqrt{\vert z''/z\vert }$ decreases from its initial value until 
inflation sets in, after which it begins to rise.
(Actually, the quantity $z''/z$ is negative during the initial kinetic 
dominated regime and turns positive during the transition to the 
inflationary epoch.
Hence, we have plotted the quantity $\sqrt{\vert z''/z\vert}$.)
It is well known that $\sqrt{z''/z} \simeq a\,H$ in slow roll inflation, as
is reflected in the linear growth of $\sqrt{z''/z}$ (in
the log-linear plot) at later times.
Interestingly, we find that  $\sqrt{\vert z''/z \vert}= {\cal O}(a\,H)$ even 
in the initial fast roll phase. The wave numbers of three modes, viz. 
$k=10^{-5}\,\mpcinv$, $10^{-4}\,\mpcinv$ and $10^{-2}\,\mpcinv$,
have also been indicated (in blue) to highlight the differences in their 
evolution.
While the first mode always remains in the super-Hubble domain (i.e. 
$k <\sqrt{\vert z^{\prime\prime}/z\vert}$), 
the second and the third modes 
spend a certain amount of time in the sub-Hubble regime (i.e. $k > 
\sqrt{\vert z^{\prime\prime}/z\vert}$) before they cross over to the 
super-Hubble regime.}\label{fig:zeebyz}
\end{figure}

\begin{figure}
\begin{center}
\includegraphics[width=15.00cm]{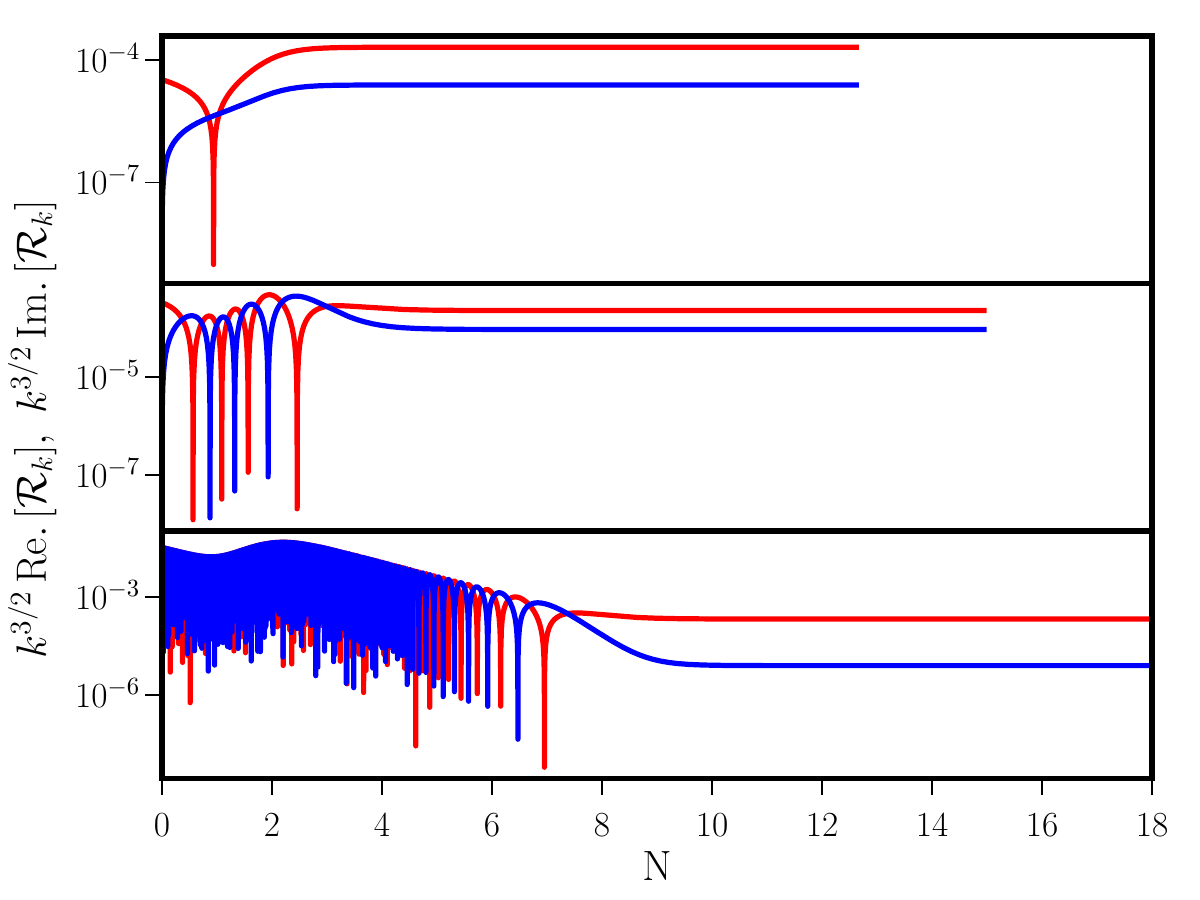}
\end{center}
\vskip -15pt
\caption{The evolution of the Fourier modes~$f_k$ of the curvature perturbation has 
been plotted as a function of $e$-folds~$N$ in a typical inflationary scenario with 
an initial epoch of kinetic domination.
In order to illustrate the oscillations, we have plotted the evolution of the
amplitudes of the real (in red) and imaginary (in blue) parts of the Fourier 
modes for three different wave numbers of cosmological interest that we had 
considered in the previous figure, viz. 
$k=10^{-5}\,\mathrm{Mpc}^{-1}$,\,$10^{-4}\, \mathrm{Mpc}^{-1}$ 
and $10^{-2}\,\mathrm{Mpc}^{-1}$
(in the top, middle and bottom panels, respectively).
Note that these plots correspond to the case of QPa wherein the modes have been 
evolved from $N=0$ (when the initial conditions are imposed on the background 
scalar field) up to a point in the super-Hubble regime, when they satisfy the 
condition $k = 10^{-5}\,\sqrt{\vert z''/z\vert} \simeq 10^{-5}\,(a\,H)$.
Evidently, the large scale mode with wave number $10^{-5}\,\mathrm{Mpc}^{-1}$,
which is always in the super-Hubble regime, barely oscillates and its amplitude 
almost remains constant (cf. top panel). 
The intermediate scale mode with wave number $10^{-4}\,\mathrm{Mpc}^{-1}$
spends a limited amount of time in the sub-Hubble regime.
It oscillates a few times before its amplitude freezes soon after leaving the 
Hubble radius (cf. middle panel).
The small scale mode with wave number $10^{-2}\,\mathrm{Mpc}^{-1}$ spends an 
adequate amount of time in the sub-Hubble regime, and it reflects the behavior 
of modes in standard slow roll inflation (cf. bottom panel).
It oscillates repeatedly in the sub-Hubble regime and settles to a constant 
amplitude on super-Hubble scales.
These differences in the behavior of the different modes of cosmological interest
lead to different amplitudes at late times and hence to features in the scalar 
power and bispectra.}\label{fig:e-modes}
\end{figure}

Recall that, in the inflationary scenario, the standard practice 
is to impose the initial conditions on the perturbations in the 
sub-Hubble limit.
However, due to the initial kinetic domination, in the scenarios
of our interest, a range of large scale modes are always outside 
the Hubble radius.
As is illustrated in Fig.~\ref{fig:zeebyz}, for the initial conditions 
for the background and the best-fit values of the parameters that we shall 
work with (when the scalar power spectra are compared with the CMB data, 
see our discussion below as well as subsection~\ref{subsec:cw-cmb-d}), we 
find that a certain range of large scale modes never satisfy the condition 
$k > \sqrt{\vert z^{\prime\prime}/z \vert}$ required for imposing the 
Bunch-Davies initial conditions.
%%%%%%%%%%%%%%%%%%%%%%%%%%%%%%%%%%%%%%%%%%%%%%%%%%%%%%%%%%%%%%%%%%%%%%%%%%%%%%%%
%%%%%%%%%%%%%%%%%%%%%%%%%%%%%%%%%%%%%%%%%%%%%%%%%%%%%%%%%%%%%%%%%%%%%%%%%%%%%%%%
We shall evolve the perturbations when the initial conditions are 
imposed at two instances in the quadratic potential~(\ref{eq:qp}) 
and the Starobinsky model~(\ref{eq:sm1}).
We shall choose to impose the Bunch-Davies conditions on the perturbations 
at the time when we begin to evolve the background (i.e. at $N=0$) and at 
the onset of inflation (i.e. at $N_1$).
For convenience, we shall refer to these cases as (QPa, QPb) and (SMIa, 
SMIb), respectively.  
In Fig.~\ref{fig:e-modes}, to illustrate the differences in the behavior 
of the various modes, we have plotted the evolution of the curvature 
perturbation for three different modes of cosmological interest in the case 
of~QPa.
%%%%%%%%%%%%%%%%%%%%%%%%%%%%%%%%%%%%%%%%%%%%%%%%%%%%%%%%%%%%%%%%%%%%%%%%%%%%%%%%
%%%%%%%%%%%%%%%%%%%%%%%%%%%%%%%%%%%%%%%%%%%%%%%%%%%%%%%%%%%%%%%%%%%%%%%%%%%%%%%%

In the case of QP, we choose the initial value of the scalar field to be
$\phi_\mathrm{i}=18.85\,\Mpl$.
As we mentioned, the initial velocity of the field is determined by the 
choice $\epsilon_{1\mathrm{i}}=2.99$.
Under these conditions, the best-fit values for the mass~$m$ of the scalar 
field prove to be $6.41\times10^{-6}\,\Mpl$ and $6.17\times10^{-6}\,\Mpl$
in the cases of QPa and QPb, respectively
(cf. Tab.~\ref{tab:chi2-bfv-recent}).
Moreover, in these cases, the best-fit values of $N_\ast$ turn out to be 
$55.06$ and $57.32$.
For the above parameter values and initial conditions, the scalar field 
rolls down the potential for about $65$ $e$-folds (counted from $N=0$ when 
the scalar field is at $\phi_\mathrm{i}$), before inflation is terminated 
close to the minimum of the quadratic potential.

In the case of SMI, we choose the initial value of the scalar field 
to be $\phi_\mathrm{i}=8.37\,\Mpl$, with $\epsilon_{1\mathrm{i}}=2.99$.
We find that the best-fit values for the parameter $\Lambda$ in the cases
of SMIa and SMIb prove to be $9.66\times10^{-10}\,\Mpl^4$ and 
$8.99\times10^{-10}\,\Mpl^4$, respectively (cf. Tab.~\ref{tab:chi2-bfv-recent}).
Also, the corresponding best-fit values for $N_\ast$ turn out to be 
$53.22$ and $55.19$.
For the above initial conditions and parameter values, we find that 
inflation ends after approximately $64$ $e$-folds.

\begin{figure}[!t]
\begin{center}
\includegraphics[width=8.50cm]{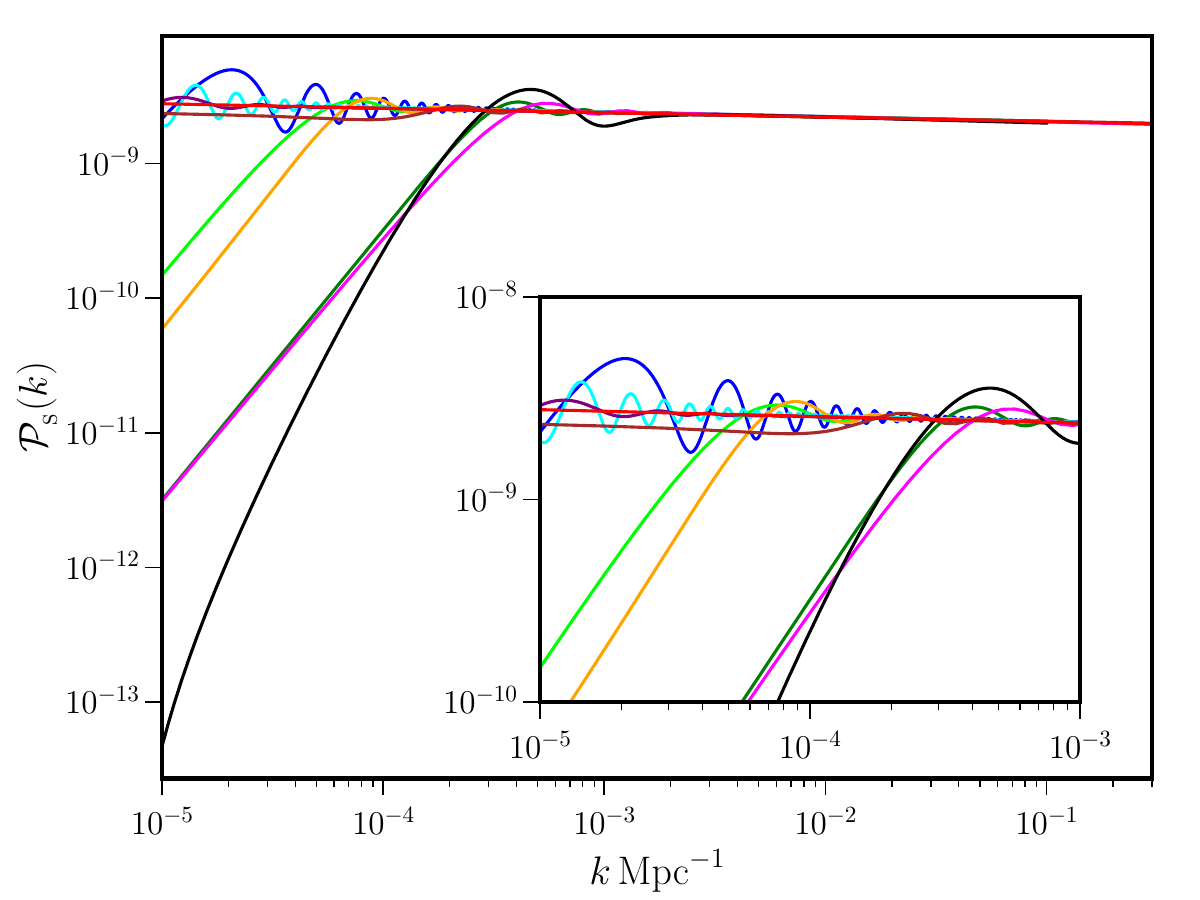}
\includegraphics[width=8.50cm]{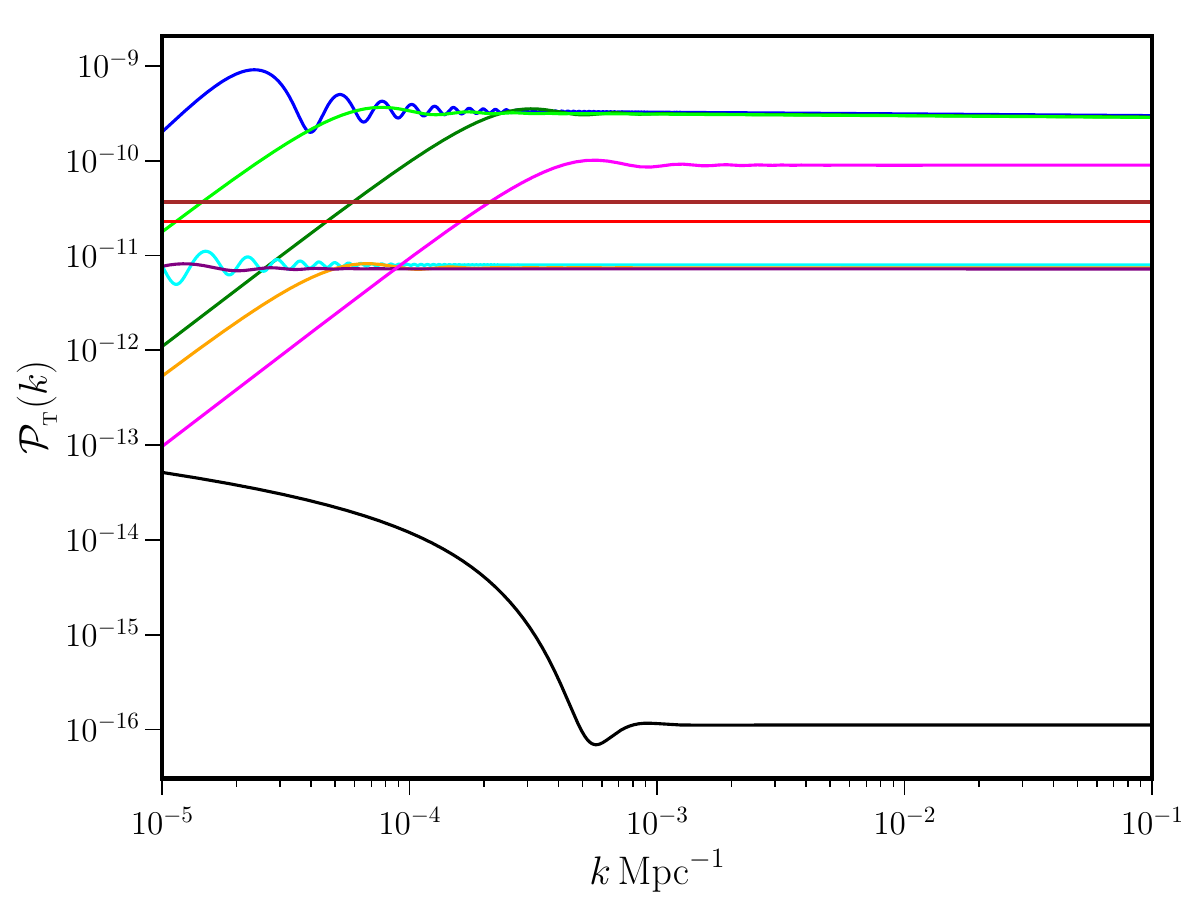}
\end{center}
\vskip -15pt
\caption{The scalar (on the left) and tensor (on the right) power spectra 
evaluated either analytically or numerically have been plotted in the various 
inflationary scenarios of our interest. 
We have plotted the best-fit power spectra in all the different models
we have considered: viz. the cases of the quadratic potential (QPa, QPb 
and QPc, in blue, green and lime) and the first Starobinsky model (SMIa, 
SMIb and SMIc, in cyan, orange and purple) with kinetically dominated 
initial conditions and their duals, the second Starobinsky model (SMII,
in brown), punctuated inflation (PI, in black) and the hard cut-off model
(HCO, in magenta).
For comparison we have also included the best-fit, 
featureless, nearly scale invariant power law spectra~(in red).
While most models exhibit a cut-off on large scales, the drop in scalar 
power is the sharpest in PI than in the other cases.
As we shall see, the scalar power spectrum in PI leads to the largest 
level of improvement in the fit to the CMB data.
Moreover, all the models lead to oscillations before the spectra turn 
nearly scale invariant and, understandably, the amplitude of the 
oscillations is the smallest in the case of HCO, since it involves 
only slow roll.
Note that, as far as the tensor power spectra are concerned, the scenarios 
involving SMI predict lower amplitudes than that of QP.
Also, the drop in tensor power on large scales in models with initial kinetic
domination, their duals and HCO, is similar to what occurs in the case of the
the scalar power spectrum.
Evidently, this behavior is due to the imposition of the standard initial 
conditions on the modes when they are in the super-Hubble regime.
Besides, we should point out that PI has the lowest tensor power amongst 
all the models of interest.}\label{fig:ps}
\end{figure}

Having evolved the background and the perturbations, we evaluate the 
power spectra at a suitably late time when {\it all}\/ the modes of 
cosmological interest (say, $10^{-5} < k< 1\,\mathrm {Mpc}^{-1}$) are 
sufficiently outside the Hubble radius.
One finds that all the scalar and tensor power spectra, generically, 
exhibit a drop in power on large scales, as illustrated in Fig.~\ref{fig:ps}. 
%%%%%%%%%%%%%%%%%%%%%%%%%%%%%%%%%%%%%%%%%%%%%%%%%%%%%%%%%%%%%%%%%%%%%%%%%%%%%%%
%%%%%%%%%%%%%%%%%%%%%%%%%%%%%%%%%%%%%%%%%%%%%%%%%%%%%%%%%%%%%%%%%%%%%%%%%%%%%%%
In fact, the suppression in power occurs when the Bunch-Davies 
initial conditions are imposed over modes that never satisfy 
the sub-Hubble condition $k > \sqrt{z''/z}$ or $k > \sqrt{a''/a}$.
Moreover, as is expected in any transition, the power spectra exhibit
a burst of oscillations before they turn nearly scale invariant on 
smaller scales.
Further, as far as the tensor power spectra are concerned, the scenarios 
involving the Starobinsky potential lead to a lower tensor power overall, 
when compared to the quadratic potential. 
However, there is a small difference in the scale at which the onset 
of the suppression occurs in the tensor power when compared to the 
scalar power in a given model. 
This can be attributed to the difference in the behavior of the 
quantities $z''/z$  and $a''/a$ that govern the evolution of the scalar 
and tensor modes respectively [cf. Eqs.~\eqref{eq:mse} and~\eqref{eq:mse-t}].
We should point out that similar scalar and tensor power spectra,
with lower power on large scales, can also be arrived at in other potentials 
that permit slow roll inflation (in this context, see App.~\ref{app:ii}).

%%%%%%%%%%%%%%%%%%%%%%%%%%%%%%%%%%%%%%%%%%%%%%%%%%%%%%%%%%%%%%%%%%%%%%%%%%%%%%%

\subsubsection{Another model due to Starobinsky}

The second scenario we shall consider is another model due to Starobinsky 
which is described by the following linear potential with an abrupt change 
in its slope~\cite{Starobinsky:1992ts,Arroja:2012ae,Martin:2011sn}:
\begin{eqnarray}
V(\phi)=\l\{\begin{array}{cc}
V_0+A_+\,\l(\phi-\phi_0\r) & \mathrm{for}~\phi>\phi_0,\\
V_0+A_-\,\l(\phi-\phi_0\r) & \mathrm{for}~\phi<\phi_0,\\
\end{array}\r.\label{eq:v-sm}
\end{eqnarray}
where $A_-\neq A_+$.
In order to distinguish from the first Starobinsky model, we shall 
refer to the scenario described by this potential as Starobinsky 
model~II (SMII, hereafter).
We should mention here that, to permit numerical analysis, one often 
works with a smoothened form of the above potential given 
by~\cite{Martin:2014kja}
\begin{eqnarray}
V(\phi) 
= V_0 + \f{1}{2}\,\l(A_{+} + A_{-}\r)\,\l(\phi-\phi_0\r) 
+ \f{1}{2}\,\l(A_{+} - A_{-}\r)\,\l(\phi-\phi_0\r)\,
\tanh{\l(\f{\phi-\phi_0}{\Delta\phi}\r)}.\label{eq:sm2}
\end{eqnarray}

It is useful here to briefly describe the dynamics that arises in the 
model.
If we work with parameters such that the constant term $V_0$ in the 
potential dominates, then the first slow roll parameter $\epsilon_1$ 
always remains fairly small through most of the evolution. 
One also finds that, in such a case, there arise two stages of slow 
roll inflation with a brief period of departure from slow roll.
The deviation from slow roll is reflected in the large values of the 
second and the third slow roll parameters, viz. $\epsilon_2$ and 
$\epsilon_3$ (with $\epsilon_{n+1}=\d \ln \epsilon_n/\d N$, for $n>1$), 
which occur briefly when the scalar field crosses~$\phi_0$.
In fact, the small value for $\epsilon_1$ permits one to express the 
scalar modes in terms of the de Sitter modes and thereby evaluate 
the power spectrum even analytically.
It can be shown that the power spectrum in the model can be expressed 
as~\cite{Starobinsky:1992ts,Arroja:2012ae,Martin:2011sn}
\begin{eqnarray}
\ps(k) 
&\simeq & 
\As\,\bigg\{1 - \f{3\,\Delta A}{A_+}\,\f{k_0}{k}\,
\l[\l(1 - \f{k_0^2}{k^2}\r)\,\mathrm{sin}\l(\f{2\,k}{k_0}\r) 
+ \f{2\,k_0}{k}\,\mathrm{cos}\l(\f{2\,k}{k_0}\r)\r]\nn\\
& & +\, \f{9\,\Delta A^2}{2\,A^2_+}\,\f{k^2_0}{k^2}\, \l(1 + \f{k_0^2}{k^2}\r)\, 
\l[1 + \f{k^2_0}{k^2}
-\f{2\,k_0}{k}\,\mathrm{sin}\l(\f{2\,k}{k_0}\r) 
+ \l(1 - \f{k_0^2}{k^2}\r)\, 
\mathrm{cos}\l(\f{2\,k}{k_0}\r)\r]\bigg\},\qquad\;\;\label{eq:ps-sm2}
\end{eqnarray}
where we have set $\As=\l[3\,H_{_{\mathrm{I}}}^3/(2\,\pi\,A_{-})\r]^2$, 
while $\Delta A = A_- - A_+$ and $H_{_{\mathrm{I}}}^2 \simeq V_0/(3\,\Mpl^2)$.
Note that, since the above analytical result for 
the scalar power spectrum has been arrived at using the de Sitter 
modes, the spectrum is strictly scale invariant on small scales.
In order to account for a tilt, while comparing with the CMB 
data, we multiply the above power spectrum by $(k/k_\ast)^{\ns-1}$. 
The tensor power spectrum is assumed to be of constant amplitude 
throughout the range of wave numbers, as the features in the model
occur only in the scalar power spectrum.
The tensor amplitude is arrived at through a constant tensor-to-scalar
ratio~$r$, which is defined through the relation $\pt = r\,\ps(k_\ast)$, 
with $k_\ast$ being the pivot scale.
As we shall discuss later, upon comparing such a spectrum with the 
CMB data, we obtain the following best-fit values for the parameters
involved:~$\As = 2.11\times 10^{-9}$, $\ns = 0.97$, 
$k_0=6.32 \times 10^{-5}\,\mpcinv$, $\Delta\,A/A_+ = -0.074$ and $r=0.017$.
We have illustrated the best-fit scalar and tensor power spectra in 
Fig.~\ref{fig:ps}.
As should be clear from the figure, the scalar power spectrum exhibits a 
step-like feature on large scales and is nearly invariant at small scales.
It should be pointed out that the height of the step in the power spectrum is
essentially determined by the difference in the slopes $A_+$ and $A_-$.

%%%%%%%%%%%%%%%%%%%%%%%%%%%%%%%%%%%%%%%%%%%%%%%%%%%%%%%%%%%%%%%%%%%%%%%%%%%%%%%

\subsubsection{The punctuated inflationary scenario}

The third scenario we shall consider is the so-called punctuated 
inflationary scenario (referred to hereafter as PI) achieved with 
the aid of the potential~\cite{Jain:2008dw,Jain:2009pm}
\begin{equation}
V(\phi) = \f{1}{2}\,m^2\,\phi^2 
- \f{2}{3}\,m^2\,\phi_0^2\,\l(\f{\phi}{\phi_0}\r)^3 
+ \frac{1}{4}\,m^2\,\phi_0^2\,\l(\frac{\phi}{\phi_0}\r)^4.
\label{eq:pi}
\end{equation}
The potential contains a point of inflection at $\phi=\phi_0$.
If one starts with a suitably large initial value of the scalar 
field such that $\phi\gg\phi_0$, the potential admits two stages 
of slow roll inflation separated by a brief departure (for less 
than an $e$-fold) from inflation.
We shall choose to work with $\phi_{\mathrm{i}}= 
12.00\,\Mp$ and $\epsilon_{1{\mathrm{i}}} = 2\times10^{-3}$.
On setting $\phi_0=1.9654\,\Mpl$, we obtain the best-fit value 
for the parameter $m$ to be $7.16\times10^{-8}\, \Mpl$.
We have plotted the resulting scalar and tensor power spectra in Fig.~\ref{fig:ps}.
Note that the scalar power spectrum exhibits a sharp drop in power on 
large scales. The model also predicts a very low amplitude for the tensor 
power throughout the range of wave numbers of interest.
However, the model has a major drawback.
One finds that, in order for the drop in power to occur at wave numbers 
roughly corresponding to the Hubble scale today, the largest scale has 
to leave the Hubble radius during inflation considerably  
(about $30$--$35$~$e$-folds) earlier
than the nominally accepted upper bound of about $65$ $e$-folds, when counted 
from the end of inflation (for a discussion on this upper bound, see
Refs.~\cite{Dodelson:2003vq,Liddle:2003as}).
For the above values of the parameters, we find that 
inflation lasts for about $110$ $e$-folds and the pivot scale itself exits 
the Hubble radius at about $90.61$ $e$-folds before the end of inflation.
Despite the drawback, we believe the model is interesting for the reason that,
amongst the different models we consider, it leads to the largest improvement
in the fit to the CMB data.
We shall briefly comment about the model further in concluding section.

%%%%%%%%%%%%%%%%%%%%%%%%%%%%%%%%%%%%%%%%%%%%%%%%%%%%%%%%%%%%%%%%%%%%%%%%%%%%%%%

\subsubsection{The hard cut-off model}\label{sec:hco}
 
It would be interesting to analytically describe the model with 
kinetically dominated initial conditions and evaluate the 
corresponding observable quantities of interest.
However, it proves to be a bit cumbersome to do so.
A simpler model, which permits complete analytical evaluation of
the scalar power and bispectra corresponds to a situation wherein 
the scalar field starts {\it on}\/ the inflationary attractor at 
some given conformal time, say, $\ei$.
We shall refer to such a scenario as the hard cut-off model (or,
simply, HCO).
The attractive aspect of the initially kinetically dominated model
is that inflation begins naturally at a specific time when the 
velocity of the scalar field decreases below a threshold value
as it rolls down the potential.
In contrast, in the hard cut-off model, we have to {\it a priori}\/ 
assume that inflation begins at a specific time with the scalar field 
being on the attractor. 

Since the model involves only slow roll, it is straightforward 
to arrive at the Fourier modes~$f_k$ describing the curvature
perturbation.
As is well known, during slow roll, the scalar mode $f_k$, in 
general, can be expressed in terms of the de Sitter solutions as
\begin{equation}
f_k(\eta) 
= \frac{i\,H_{_{\rm I}}}{\Mpl\,\sqrt{4\,k^3\,\epsilon_1}}\,
\bigl[\alpha_k\, (1+i\,k\,\eta)\,{\rm e}^{-i\,k\,\eta}
- \beta_k\,(1-i\,k\,\eta)\,{\rm e}^{i\,k\,\eta}\bigr],\label{eq:fk-hco}
\end{equation}
where $H_{_{\rm I}}$ represents the Hubble scale during inflation and 
$\epsilon_1$ denotes the first slow roll parameter.
The quantities $\alpha_k$ and $\beta_k$ are the so-called Bogoliubov 
coefficients.
If one imposes the standard Bunch-Davies initial conditions in the 
sub-Hubble limit, then one will have $\alpha_k=1$ and $\beta_k=0$.
In our case, we shall impose the initial conditions at the time $\ei$
irrespective of whether the modes are inside or outside the Hubble radius.
In such a case, we obtain the Bogoliubov coefficients $\alpha_k$ and 
$\beta_k$ to be
\begin{subequations}
\begin{eqnarray}
\alpha_k &=& 1+\f{i}{k\,\ei}-\frac{1}{2\,k^2\,\ei^2}
=1-\f{i\,\ki}{k}-\f{\ki^2}{2\,k^2},\\
\beta_k &=& -\f{1}{2\,k^2\,\ei^2}\,{\rm e}^{-2\,i\,k\,\ei}
=-\f{\ki^2}{2\,k^2}\,{\rm e}^{2\,i\,k/\ki},
\end{eqnarray}
\end{subequations}
where we have set $\ki=-1/\ei$.
Note that, as $\ei\to-\infty$ (i.e. as $\ki\to 0$), $\alpha_k\to 1$
and $\beta_k\to 0$, which corresponds to the conventional sub-Hubble,
Bunch-Davies initial conditions often imposed on all the modes. 

With the modes $f_k$ at hand, it is now straightforward to evaluate 
the resulting power spectrum by substituting the modes in the 
expression~(\ref{eq:ps}) and taking the late time (i.e. $\eta\to 0$)
limit.
One can easily show that the power spectrum can be written as
\begin{equation}
\ps(k)
= \As\,\vert \alpha_k - \beta_k \vert ^2
= \As\, \biggl[1 +\f{\ki^4}{2\,k^4}-\f{\ki^3}{k^3}\, \sin \l(\f{2\,k}{\ki}\r)
+\left(\f{\ki^2}{k^2} -\frac{\ki^4}{2\,k^4}\right)\, 
\cos \l(\frac{2\,k}{\ki}\r)\biggr],\quad\;\;\label{eq:ps-hco}
\end{equation}
where we have set $\As=H_{_{\mathrm{I}}}^2/(8\,\pi^2\,\epsilon_1)$.
We find that this analytical expression matches the corresponding numerical
result very well, say, in QP or SMI, modulo at small scales where the de 
Sitter modes are not adequate to capture the spectral tilt that arises in a
realistic model.
Therefore, when comparing the HCO model with the CMB data, 
to allow for the spectral tilt at small scales, we have multiplied the above 
scalar power spectrum by~$(k/k_\ast)^{\ns-1}$ as in the case of SMII.
Moreover, we have defined the tensor power spectrum to be $\pt(k) = r\,\ps(k)$,
with the tensor-to-scalar ration~$r$ assumed to be a constant.
Such a scale-dependent definition (in contrast to the SMII case, where the 
tensor power spectrum was scale invariant) is important because the model 
predicts features in the tensor power spectrum similar to that in the scalar 
power spectrum, and we have consistently accounted for it in the analysis.
We obtain the following best-fit values for the parameters 
involved:~$\As=2.09\times 10^{-9}$, $\ns = 0.96$, 
$\ki = 2.03\times 10^{-4}\,\mpcinv$, and $r=0.043$.
In Fig.~\ref{fig:ps}, we have plotted the analytical scalar
and tensor power spectra, with the~$(k/k_\ast)^{\ns-1}$ term included, 
corresponding to the above mentioned best-fit values.
Clearly, the power spectra exhibit a suppression of power on large scales
as in the case of the other models.
As we shall discuss later, the HCO model allows us to evaluate the scalar 
bispectrum too analytically.
The analytical calculations prove to be handy as they permit us to test
the numerical results against the analytical results in a situation wherein 
the Bunch-Davies initial conditions are imposed on super-Hubble scales.

%%%%%%%%%%%%%%%%%%%%%%%%%%%%%%%%%%%%%%%%%%%%%%%%%%%%%%%%%%%%%%%%%%%%%%%%

\subsubsection{A dual to initial kinetic domination}

We shall now discuss a situation which we shall refer to as the dual 
to the scenario with kinetically dominated initial conditions.
Recall that, in the model with initial kinetic domination, the 
scalar field starts with a large velocity.
Evidently, this corresponds to a situation wherein the field begins
from a point away from the inflationary attractor. 
It is interesting to examine the effects on the power spectrum in a
scenario with a finite duration of inflation where the field starts 
with a small velocity (than its value on the attractor) rather than 
a large velocity.
As in the hard cut-off model, there is no natural way of terminating
inflation (when one goes back in time) in such a case.
Therefore, we shall assume that inflation begins at a specific time 
and that the Bunch-Davies initial conditions are imposed on super-Hubble 
scales for a range of modes.
A version of such a scenario has been considered previously in the 
literature and we find that they are referred to as non-attractor 
models of inflation (in this context, see, for instance,
Ref.~\cite{Cai:2016ngx,Cai:2017bxr}).
Under these conditions, we find that, as the field evolves towards 
the attractor, there occurs a sharp drop in power on large scales 
and a regime of oscillations arises over intermediate scales before 
the spectrum turns nearly scale invariant on small scales. 
We shall refer to this case as QPc and SMIc when implemented in the 
quadratic potential~(\ref{eq:qp}) and Starobinsky model~(\ref{eq:sm1}), 
respectively.
In the case of QPc, we choose $\phi_\mathrm{i} = 16.00\,\Mpl$ and 
$\epsilon_{1\mathrm{i}} = 10^{-4}$, which leads to inflation of 
about $65$ $e$-folds.
In the case of SMIc, we work with $\phi_\mathrm{i} = 5.52\,\Mpl$ and 
$\epsilon_{1\mathrm{i}} = 10^{-4}$, which too results in inflation 
lasting for about $65$ $e$-folds.
We find that the best-fit values for the parameter in these models 
prove to be $m=6.15\times10^{-6}\,\Mpl$ and 
$V_0 = 8.75\times 10^{-10}\,\Mpl^4$.
The pivot scale exits the Hubble radius at $57.32$ and $55.87$ $e$-folds 
before the end of inflation in the cases of QPc and SMIc, respectively.
In Fig.~\ref{fig:ps}, we have illustrated the power spectra (that 
lead to the best-fit to the CMB data) in the dual scenario, viz. QPc 
and SMIc, along with the spectra arising in the cases with initial 
kinetic domination, i.e. QPa, SMIa, QPb and SMIb  (as well as the 
other models of interest).
Clearly, the kinetically dominated model and its dual generate spectra 
with roughly similar features.
We find that the drop in power at large scales have the same shape in 
both the scenarios and is mostly independent of the initial velocity
of the field.

%%%%%%%%%%%%%%%%%%%%%%%%%%%%%%%%%%%%%%%%%%%%%%%%%%%%%%%%%%%%%%%%%%%%%%%%%%%%%%

\subsection{Performance against the CMB data}\label{subsec:cw-cmb-d}

We have compared all the models we have described in the previous subsection 
against the recent Planck data~\cite{Aghanim:2019ame}.
In this subsection, we shall discuss the assumptions we have made while 
comparing the models with the CMB data, the priors on the parameters we
have worked with and present the final results for the CMB angular power
spectrum.

We have taken into account both the scalar and tensor power spectra arising 
in the models of interest when comparing against the CMB data.
We have modified the CAMB package suitably to include 
the scalar power spectra arising in the models of our 
interest~\cite{Lewis:1999bs}. 
We make use of CosmoMC to carry out the comparison of the models with the 
CMB data and arrive at the respective likelihoods~\cite{Lewis:2002ah}. 
We have worked with the 2018 release of Planck data, which comprises of the 
likelihoods of the TT, TE as well as the EE correlations,
along with the lensing likelihood~\cite{Aghanim:2019ame}. 
We have included nonlinear lensing in the calculation of the CMB angular power 
spectra which has a significant effect over small scales.
For models with inflationary spectra calculated numerically, we have evaluated 
the power spectra at $2000$ points over the following range of wave 
numbers:~$10^{-6}\leq k\leq  10\,\mpcinv$. 
We should mention that, in CAMB, the maximum value of the multipole $\ell$
is set to be $2700$ to compute the CMB angular power spectrum~$C_\ell$.
We perform an MCMC sampling of the posterior distribution of the parameter
space using CosmoMC for each model and arrive at the best-fit~$\chi^2$ and
the corresponding set of parameter values using the in-built package 
called GetDist.

In Tab.~\ref{tab:bp}, we have listed the priors on the four background
cosmological parameters that we have worked with.
%%%%%%%%%%%%%%%%%%%%%%%%%%%%%%%%%%%%%%%%%%%%%%%%%%%%%%%%%%%%%%%%%%%%%%%%%%%%%%%
\begin{table}[!t]
\centering
\begin{tabular}{|c|c|c|}
\hline
Parameter & Lower limit & Upper limit\\
\hline
$\Omega_\mathrm{b}\,h^2$ & $0.005$ & $0.1$\\ 
\hline
$\Omega_\mathrm{c}\,h^2$ & $0.001$ & $0.99$\\ 
\hline
$\theta$ & $0.5$ & $10$\\ 
\hline
$\tau$ & $0.01$ & $0.8$\\ 
\hline
\end{tabular}
\caption{The background cosmological parameters which we have varied and the 
priors that we have worked with.
These are the standard set of background parameters that are often considered 
while comparing with the CMB data~\cite{Aghanim:2018eyx}.}\label{tab:bp}
\end{table}
%%%%%%%%%%%%%%%%%%%%%%%%%%%%%%%%%%%%%%%%%%%%%%%%%%%%%%%%%%%%%%%%%%%%%%%%%%%%%%%
In Tab.~\ref{tab:ip}, we have listed the priors on the parameters 
describing the various inflationary models of our interest. 
Previous experience suggests that the drop in power leading to an
improved fit to the CMB data is expected to occur around the wave number 
$k\simeq 10^{-4}\,\mpcinv$. 
Hence, while choosing the range of priors for the model parameters which determine
the location of the drop in the scalar power (such as $N_\ast$), we have made sure 
that the feature occurs over the wave numbers $10^{-5}\lesssim k \lesssim 10^{-3}\,
\mpcinv$.
%%%%%%%%%%%%%%%%%%%%%%%%%%%%%%%%%%%%%%%%%%%%%%%%%%%%%%%%%%%%%%%%%%%%%%%%%%%%%%%
%%%%%%%%%%%%%%%%%%%%%%%%%%%%%%%%%%%%%%%%%%%%%%%%%%%%%%%%%%%%%%%%%%%%%%%%%%%%%%%
\begin{table}
\centering
\begin{tabular}{|c|c|c|c|c|c|}
\hline
Model & $\mathrm{ln}\, (\As\times10^{10})$ 
& $\ns$ & $r$ & ~~~~$\mathrm{log}_{10}(\ki/\mathrm{Mpc}^{-1})$ & $\Delta A/A_+$ \\
& & & & or~$\,\mathrm{log}_{10}(k_0/\mathrm{Mpc}^{-1})$ &\\
\hline
PL & $[1.61, 3.91]$ & $[0.8, 1.2]$ & $[0,2]$ & - &  - \\
\hline
HCO & $[1.61, 3.91]$ & $[0.8, 1.2]$ & $[0,2]$ & $[-4,-2]$ & - \\
\hline
SMII & $[1.61, 3.91]$ & $[0.8, 1.2]$ & $[0,2]$ & $[-5,-3]$ & $[-0.999,0.700]$\\
\hline
\end{tabular}
%%%%%%%%%%%%%%%%%%%%%%%%%%%%%%%%%%%%%%%%%%%%%%%%%%%%%%%%%%%%%%%%%%%%%%%%%%%%%%%
\vskip 10pt
%%%%%%%%%%%%%%%%%%%%%%%%%%%%%%%%%%%%%%%%%%%%%%%%%%%%%%%%%%%%%%%%%%%%%%%%%%%%%%%
\begin{tabular}{|c|c|c|c|c|}
\hline
Model & $N_\ast$ & $\mathrm{log}_{10}(10^{10}\,m^2/{\Mpl^2})$ 
& $\mathrm{log}_{10}(10^{10}\,\Lambda/{\Mpl^4})$\\
\hline
QPa, QPb, QPc &$[50, 60]$ &  $[-0.55, -0.25]$ & - \\
\hline
SMIa, SMIb, SMIc & $[50, 60]$ & - & $[0.8,1.2]$\\
\hline
PI &  $[87,93]$ & $[-5.20, -3.47]$ & -\\
\hline
\end{tabular}
\caption{The parameters associated with the different inflationary models
of our interest and the priors that we have worked with. 
The first set (on top) corresponds to models wherein we have made use of
the analytical results for the power spectra and the second set (at the 
bottom) corresponds to models wherein we have evaluated the spectra
numerically.}\label{tab:ip}
\end{table}
%%%%%%%%%%%%%%%%%%%%%%%%%%%%%%%%%%%%%%%%%%%%%%%%%%%%%%%%%%%%%%%%%%%%%%%%%%%%%%%
%%%%%%%%%%%%%%%%%%%%%%%%%%%%%%%%%%%%%%%%%%%%%%%%%%%%%%%%%%%%%%%%%%%%%%%%%%%%%%%
\begin{table}[!t]
\centering
\begin{tabular}{|c|c|c|c|c|c|c|}
\hline
Model & $\mathrm{ln}\,(\As\times10^{10})$ & $\ns$ & $r$
& $\mathrm{log}_{10}(\ki {\,(\mathrm{or})\,} k_0/{\rm Mpc}^{-1})$ 
& $(A_--A_+)/A_+$ & $\Delta \chi^2$\\
\hline
PL & $3.042$ & $0.967$ & $0.011$ & - &  - & - \\
\hline
HCO & $3.039$ & $0.962$ & $0.043$ & $-3.692$ & - & $-0.096$\\
\hline
SMII & $3.048$ & $0.969$ & $0.017$ & $-4.199$ & $-0.074$ & $-0.672$\\
\hline
\end{tabular}
%%%%%%%%%%%%%%%%%%%%%%%%%%%%%%%%%%%%%%%%%%%%%%%%%%%%%%%%%%%%%%%%%%%%%%%%%%%%%%%
\vskip 10pt
%%%%%%%%%%%%%%%%%%%%%%%%%%%%%%%%%%%%%%%%%%%%%%%%%%%%%%%%%%%%%%%%%%%%%%%%%%%%%%%
\begin{tabular}{|c|c|c|c|c|c|}
\hline
Model & $N_\ast$ & log$_{10}(10^{10}m^2/{\Mpl^2})$ 
& log$_{10}(10^{10}\Lambda/{\Mpl^4})$ & $\Delta \chi^2$\\
\hline
QPa & $55.06$ & $-0.386$ & - & $2.494$\\
\hline
QPb & $57.32$ & $-0.419$ & - & $-0.384$\\
\hline
QPc & $57.32$ & $-0.422$ & - & $-0.014$\\
\hline
SMIa & $53.22$ & - & $0.985$ & $-0.896$\\
\hline
SMIb & $55.19$ & - & $0.954$ & $-0.880$\\
\hline
SMIc & $55.87$ & - & $0.942$ & $-1.170$\\
\hline
PI & $90.61$ & $-4.290$ & - & $-1.746$\\
\hline
\end{tabular}
\caption{The best-fit values of the inflationary parameters and the extent of 
improvement in $\chi^2$ with respect to the standard power law case, arrived at 
by comparing the models with the recent CMB data.
As in the previous table, the first set (on top) corresponds to models with 
analytical forms for the scalar power spectra and the second set (at the bottom)
corresponds to those cases wherein the spectra have been evaluated numerically.
We have defined $\Delta \chi^2=(\chi^2_\mathrm{model}-\chi^2_{_{\mathrm{PL}}})$
so that a negative value for $\Delta \chi^2$ implies an improvement in the 
fit with respect to the PL case.
Note that PI leads to the largest improvement in the fit to the data.
The dataset we have used is the following combination of likelihoods:~TT 
+ TE + EE + low~$\ell$ + low~E + lensing, from the Planck 2018 data 
release.
The models were compared while accounting for tensors as well as non-linear 
lensing. 
We should mention that the best-fit value for $\chi^2$ in the PL case
we have obtained is $\chi^2 = 2\,(1390.928)=2781.856$.
The corresponding value quoted by the Planck team is 
$\chi^2 = 2\,(1391.104) = 2782.208$, which is close to the value we 
obtain~\cite{Aghanim:2018eyx}.}\label{tab:chi2-bfv-recent}
\end{table}
%%%%%%%%%%%%%%%%%%%%%%%%%%%%%%%%%%%%%%%%%%%%%%%%%%%%%%%%%%%%%%%%%%%%%%%%%%%%%%%

\begin{figure}[!t]
\centering
\includegraphics[width=15.00cm]{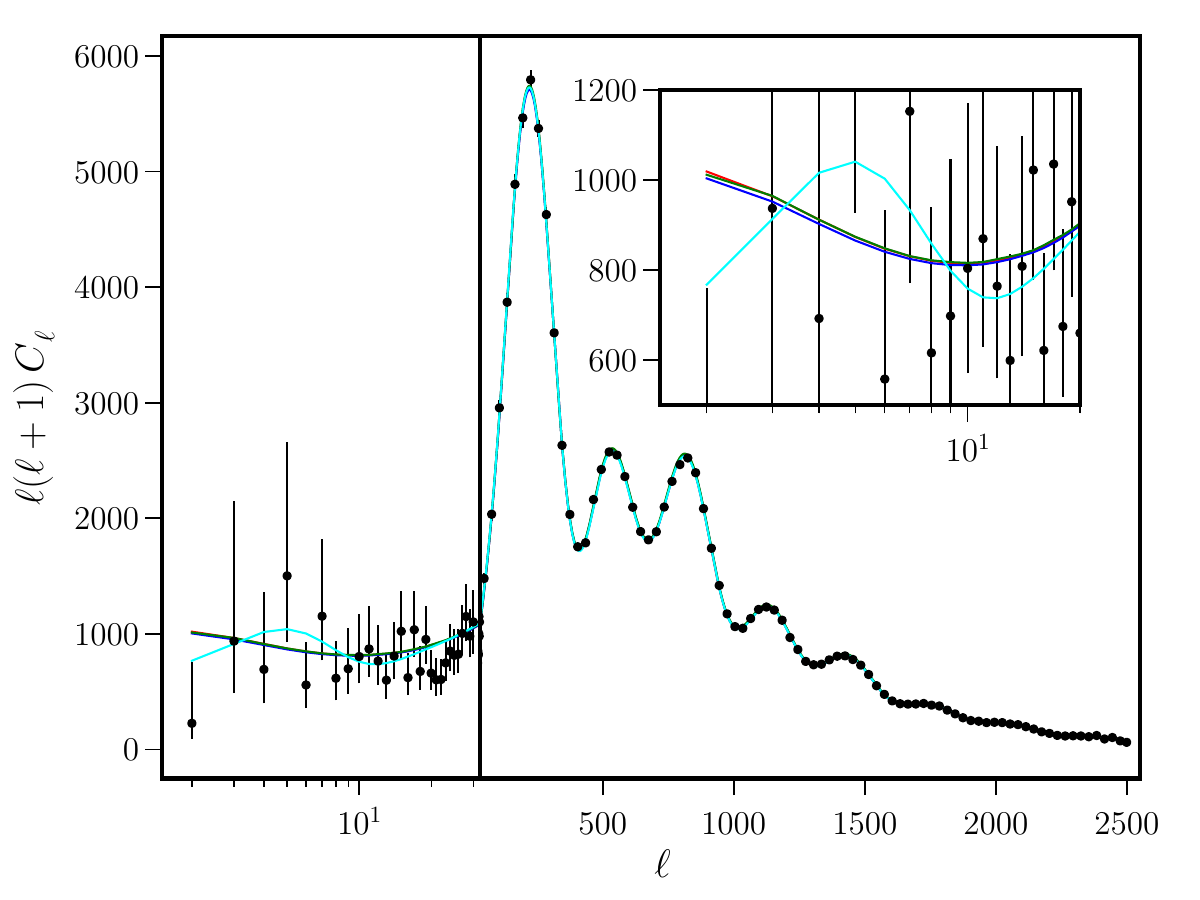}
\caption
{The best-fit CMB angular power spectra 
have been plotted for the three models SMIc (in blue), SMII (in green) and PI (in
cyan) which lead to an improvement in $\Delta\chi^2$ of $0.6\mbox{--}1.7$ in the 
fit to the recent Planck data.
In order to highlight the differences, we have also plotted the best-fit 
angular power spectrum for the PL case (in red).
We have also included the Planck 2018 data points along with their error bars 
(in black).
Note that the multipoles $\ell$ appear on a log scale until $\ell=32$ (indicated
by the vertical line) and on a linear scale for $\ell>32$.
We should point out the fact that the CMB angular spectrum in the case of 
PI exhibits an oscillation over the lower multipoles before it merges with, 
say, the result for the power law case, at the higher multipoles.
The angular spectra for the models of SMIc and SMII are suppressed to a far 
less extent when compared to the spectrum of PI over the low multipoles (as 
highlighted in the inset).}\label{fig:bf-cls}
\end{figure}

In Tab.~\ref{tab:chi2-bfv-recent}, we have listed the improvement in the 
$\chi^2$ and the best-fit values for the inflationary parameters for
the different models we have considered.
Recall that the power law~(PL) case corresponds to the simplest situation
wherein the scalar power spectrum is expressed 
as~$\ps(k)=\As\,(k/k_\ast)^{\ns-1}$.
For the PL case, the $\chi^2$ we obtain from the GetDist package is $2781.856$, 
while the value quoted by the Planck team is~$2782.208$\footnote{See the 
Planck Legacy Archive located at the following 
URL: \href{http://pla.esac.esa.int/pla/\#cosmology}
{http://pla.esac.esa.int/pla/\#cosmology}.}.
Note that the quantity $\Delta \chi^2$ is the difference in $\chi^2$ between
a given inflationary model and the PL case, with a negative value indicating 
an improvement in the fit to the data.
Evidently, PI leads to the largest improvement in the fit to the data.
Earlier, in Fig.~\ref{fig:ps}, we had plotted the inflationary scalar 
power spectra for parameter values of the various models that lead to 
the best fit to the CMB data.
In Fig.~\ref{fig:bf-cls}, we have plotted the corresponding CMB angular 
power spectra for three of the models which lead to reasonable levels of 
improvement (with $\Delta \chi^2\simeq 0.6\mbox{--}1.7$) in their fit to the data.
%%%%%%%%%%%%%%%%%%%%%%%%%%%%%%%%%%%%%%%%%%%%%%%%%%%%%%%%%%%%%%%%%%%%%%%%%%%%%%%
%%%%%%%%%%%%%%%%%%%%%%%%%%%%%%%%%%%%%%%%%%%%%%%%%%%%%%%%%%%%%%%%%%%%%%%%%%%%%%
Lastly, we should mention that, in App.~\ref{app:ccp}, we have illustrated 
the marginalized posterior distribution on the inflationary parameters in
the models SMIc, PI and SMII.
We have also illustrated the constraints on the Hubble parameter $H_0$ 
that one obtains in these cases.
Interestingly, we find that, though PI modestly improves the fit to the 
CMB data, it seems to exacerbate the so-called Hubble tension, a topic 
that is considerable interest today~\cite{DiValentino:2021izs,Freedman:2021ahq}.

Having described the alternative scenarios resulting in scalar spectra
with a sharp drop in power on large scales, let us now turn to the 
evaluation of the scalar non-Gaussianities in these models. 

%%%%%%%%%%%%%%%%%%%%%%%%%%%%%%%%%%%%%%%%%%%%%%%%%%%%%%%%%%%%%%%%%%%%%%%%%%%%%%%

\section{The third order action and the surface terms}\label{sec:toa}

In order to evaluate the scalar bispectrum, one requires the action 
describing the curvature perturbation at the third order. 
It can be shown that, at the third order, the action governing the
curvature perturbation $\cR$ can be expressed as (see, for instance, 
Refs.~\cite{Maldacena:2002vr,Seery:2005wm,Martin:2011sn,Arroja:2011yj})
\begin{eqnarray}
\delta S_3[\cR] &=& \Mpl^2\, \int_{\ei}^{\ee}\, \d\eta\, \int \d^3{\bm x}\, 
\biggl[a^2\, \epsilon_1^2\, \cR\,\cR'^2
+ a^2\, \epsilon_1^2\, \cR\,(\pa\cR)^2
- 2\,a\,\epsilon_1\, \cR^{\prime}\, (\pa\cR)\,(\partial\chi)\nn\\ 
& &+\, \f{a^2}{2}\,\epsilon_1\,\epsilon_2'\,\cR^2\,\cR'
+ \frac{\epsilon_1}{2}\,(\pa\cR)\, (\pa\chi)\, \partial^2\chi
+ \frac{\epsilon_1}{4}\,\pa^2\cR\,(\partial\chi)^2 
+ 2\,{\cal F}(\cR)\, \frac{\delta {\cal L}_2}{\delta \cR}\biggr], 
\label{eq:S3}
\end{eqnarray}
where, as we have mentioned earlier, $\epsilon_2=\d\, {\rm ln}\,\epsilon_1/\d N$ 
is the second slow roll parameter, while $\pa^2\chi= a\,\epsilon_1\,\cR'$.
The quantity ${\cal F}$($\cR$) is given by
\begin{eqnarray}
{\cal F}(\cR) &=& \frac{\epsilon_2}{4}\,\cR^2 
+ \f{1}{a\,H}\,\cR\,\cR'
+ \f{1}{4\,a^2\,H^2} \biggl\{-(\pa\cR)\, (\pa\cR)
+ \pa^{-2}[\pa_i\,\pa_j\,(\pa_i\cR\,\pa_j\cR)]\biggr\} \nn \\
& & +\, \f{1}{2\,a^2\,H} \biggl\{(\pa\cR)\, (\pa\chi) 
- \partial^{-2}[\pa_i\,\partial_j\,(\pa_i\cR\,\pa_j\chi)]\biggr\}
\end{eqnarray}
and ${\cal L}_2$ denotes the Lagrangian density associated with the action
governing the curvature perturbation at the second order.
Note that $\ei$ is the conformal time when the initial conditions are imposed
on the perturbations and $\ee$ is the conformal time close to the end of 
inflation, when the power and bispectra are evaluated.
Typically, in analytical calculations, one assumes that $\ei\to-\infty$ and 
$\ee\to 0^-$.

The third order action~(\ref{eq:S3}) is arrived at from the original action 
governing the system of the gravitational and scalar fields.
A set of temporal and spatial boundary terms are often ignored in arriving at
the above action~\cite{Maldacena:2002vr,Seery:2005wm,Arroja:2011yj}.
The spatial boundary terms do not contribute to the scalar bispectrum under any
condition.
However, in cases such as the scenario involving inflation of a finite duration, 
one finds that the temporal boundary terms can contribute non-trivially.
These temporal boundary terms are given by~\cite{Arroja:2011yj}
\begin{eqnarray}
\delta S^{\rm B}_3[\cR]
&=& \Mpl^2\, \int_{\ei}^{\ee}\d\eta\,\int\d^3{\bm x}\;
\f{\d}{\d\eta}\biggl\{-9\,a^3H\,\cR^3
+\f{a}{H}\,(1-\epsilon_1)\,\cR\,(\pa\cR)^2
- \f{1}{4\,a\,H^3}\,(\pa\cR)^2\,\pa^2\cR\nn\\ 
& &- \f{a\,\epsilon_1}{H}\,\cR\,\cR'^2
-\,\f{a\,\epsilon_2}{2}\,\cR^2\,\pa^2\chi
+ \frac{1}{2\,a\,H^2}\,\cR\,\l(\pa_i\pa_j\cR\,\pa_i\pa_j\chi 
- \partial^2\cR\,\partial^2\chi\r)\nn\\
& &-\, \frac{1}{2\,aH}\,\cR\,
\l[\pa_i\pa_j\chi\,\pa_i\pa_j\chi - (\pa^2\chi)^2\r]\biggr\}. 
\label{eq:S3B}
\end{eqnarray}
It should be mentioned here that, in standard slow roll inflation, apart from 
the term involving~$\epsilon_2$, none of the above terms contribute either at 
early or at late times.
The term involving $\epsilon_2$ contributes non-trivially at late times, 
and this contribution is often absorbed through a field redefinition (in 
this context, see, for example, Refs.~\cite{Maldacena:2002vr,Arroja:2011yj}).
However, it is important to clarify that, in this work, we do not carry out 
any field redefinition.
We shall explicitly calculate all the contributions due to the bulk and the 
boundary terms~(\ref{eq:S3}) and~(\ref{eq:S3B}).

%%%%%%%%%%%%%%%%%%%%%%%%%%%%%%%%%%%%%%%%%%%%%%%%%%%%%%%%%%%%%%%%%%%%%%%%%%%%%%%

\section{Evaluating the scalar bispectrum}\label{sec:sbs}

In this section, we shall describe the numerical evaluation of the scalar 
bispectrum and the corresponding non-Gaussianity parameter $\fnl$ in the
different models of our interest.
In fact, these quantities have been calculated earlier in the cases of 
the second Starobinsky model and punctuated inflation (in this context, 
see, for example, Refs.~\cite{Martin:2011sn,Hazra:2012yn,Sreenath:2014nca}).
We should mention here that, in an earlier work, we had briefly presented the 
main results for the scalar bispectrum in models with kinetic dominated 
initial conditions~\cite{Ragavendra:2019mek}.
These models wherein the initial conditions are imposed on super-Hubble 
scales pose certain challenges and it is instructive to compare the 
numerical procedure for the computation of the bispectrum and the 
non-Gaussianity parameter $\fnl$ in the different cases.

%%%%%%%%%%%%%%%%%%%%%%%%%%%%%%%%%%%%%%%%%%%%%%%%%%%%%%%%%%%%%%%%%%%%%%%%%%%%%%%%

\subsection{The scalar bispectrum and the non-Gaussianity parameter}

Let us begin by recalling a few essential points regarding the scalar 
bispectrum $G(\vka,\vkb,\vkc)$ and the corresponding non-Gaussianity 
parameter~$\fnl(\vka,\vkb,\vkc)$, where the three wavevectors $\vka$, 
$\vkb$ and $\vkc$ form the edges of a triangle.  
In the single field inflationary scenarios of our interest, the scalar 
bispectrum is essentially the three-point function of the curvature 
perturbation in Fourier space.
The bispectrum can be arrived at by using the third order action describing 
the curvature perturbation we discussed in the previous section and the 
standard rules of perturbative quantum field theory~\cite{Maldacena:2002vr,
Seery:2005wm,Martin:2011sn,Arroja:2011yj}.

It can be shown that the scalar bispectrum can be expressed 
as (see, for instance, Refs.~\cite{Martin:2011sn,Hazra:2012yn})
\begin{eqnarray}
G(\vka,\vkb,\vkc) 
&=& \sum_{C=1}^{9}\; G_{_{C}}(\vka,\vkb,\vkc)\nn\\
&=& \Mp^2\; \sum_{C=1}^{6}\; 
\Biggl[f_{k_1}(\ee)\, f_{k_2}(\ee)\,f_{k_3}(\ee)\,
\cG_{_{C}}(\vka,\vkb,\vkc)+{\rm complex\;conjugate}\Biggr]\nn\\
& & +\, G_{7}(\vka,\vkb,\vkc) +\, G_{8}(\vka,\vkb,\vkc)
+ G_{9}(\vka,\vkb,\vkc),\label{eq:sbs}
\end{eqnarray}
where, as we mentioned earlier, $f_k$ are the Fourier modes of the 
curvature perturbation [cf. Eq.~(\ref{eq:fk-vk})], while $\ee$ 
denotes the conformal time close to the end of inflation.
The quantities $\cG_{_{C}}(\vka,\vkb,\vkc)$ represent six integrals that 
involve the scale factor, the slow roll parameters, the modes~$f_k$ and 
their time derivatives~$f_k'$.
They correspond to the six bulk terms appearing in the cubic order 
action~(\ref{eq:S3}) and are described by the following expressions: 
\begin{subequations}\label{eq:cG}
\begin{eqnarray}
\cG_1(\vka,\vkb,\vkc)
&=& 2\,i\,\int_{\ei}^{\ee} \d\eta\; a^2\, 
\epsilon_{1}^2\, \biggl(f_{k_1}^{\ast}\,f_{k_2}'^{\ast}\,
f_{k_3}'^{\ast}
+{\rm two~permutations}\biggr),\label{eq:cG1}\\
\cG_2(\vka,\vkb,\vkc)
&=&-2\,i\;\l(\vka\cdot \vkb +\,{\rm two~permutations}\r)\, 
\int_{\ei}^{\ee} \d\eta\; a^2\, 
\epsilon_{1}^2\, f_{k_1}^{\ast}\,f_{k_2}^{\ast}\,
f_{k_3}^{\ast},\label{eq:cG2}\\
\cG_3(\vka,\vkb,\vkc)
&=&-2\,i\,\int_{\ei}^{\ee} \d\eta\; a^2\,\epsilon_{1}^2\,
\biggl(\f{\vka\cdot\vkb}{k_2^2}\,
f_{k_1}^{\ast}\,f_{k_2}'^{\ast}\, f_{k_3}'^{\ast}
+{\rm five~permutations}\biggr),\label{eq:cG3}\\
\cG_4(\vka,\vkb,\vkc)
&=& i\,\int_{\ei}^{\ee} \d\eta\; a^2\,\epsilon_{1}\,\epsilon_{2}'\, 
\biggl(f_{k_1}^{\ast}\,f_{k_2}^{\ast}\,f_{k_3}'^{\ast}
+ {\rm two~permutations}\biggr),\label{eq:cG4}\\
\cG_5(\vka,\vkb,\vkc)
&=&\frac{i}{2}\,\int_{\ei}^{\ee} \d\eta\; 
a^2\, \epsilon_{1}^{3}\;\biggl(\f{\vka\cdot\vkb}{k_2^2}\,
f_{k_1}^{\ast}\,f_{k_2}'^{\ast}\, f_{k_3}'^{\ast} 
+ {\rm five~permutations}\biggr),\label{eq:cG5}\\
\cG_6(\vka,\vkb,\vkc) 
&=&\frac{i}{2}\,\int_{\ei}^{\ee}\d\eta\, a^2\, \epsilon_{1}^{3}\,
\biggl(\f{k_1^2\,\l(\vkb\cdot\vkc\r)}{k_2^2\,k_3^2}\, 
f_{k_1}^{\ast}\, f_{k_2}'^{\ast}\, f_{k_3}'^{\ast} 
+ {\rm two~permutations}\biggr).\label{eq:cG6}
\end{eqnarray}
\end{subequations}
These integrals are to be evaluated from a sufficiently early time ($\ei$), 
when the modes are typically well inside the Hubble radius, until very late 
times, which can be conveniently chosen to be a time close to the end of 
inflation~($\ee$).
We should mention here that the last term in action~(\ref{eq:S3}) involving
${\cal F}(\cR)\, ({\delta {\cal L}_2}/{\delta \cR})$ actually vanishes when 
we assume that the curvature perturbation satisfies the linear equation of 
motion [cf. Eqs.~(\ref{eq:fk-vk}) and~(\ref{eq:mse})].

In the expression~(\ref{eq:sbs}) for the scalar bispectrum, the 
terms $G_{7}(\vka,\vkb,\vkc)$, $G_{8}(\vka,\vkb,\vkc)$ and 
$G_{9}(\vka,\vkb,\vkc)$ are the contributions that arise due to 
the boundary terms~(\ref{eq:S3B}) associated with the third order
action governing the curvature perturbation.
The contribution $G_{7}(\vka,\vkb,\vkc)$ is due to the term 
containing $\epsilon_2$ in the boundary terms~(\ref{eq:S3B})
and it can be expressed as
\begin{eqnarray}
G_{7}(\vka,\vkb,\vkc)
&=& -i\,\Mpl^2\,(f_{k_1}(\ee)\,f_{k_2}(\ee)\,f_{k_3}(\ee)) \nn \\
& &\times\, \biggl[a^2\epsilon_1\epsilon_{2}\,
f_{k_1}^{\ast}(\eta)\,f_{k_2}^{\ast}(\eta)\,f_{k_3}'^{\ast}(\eta) 
+ {\rm two~permutations} \biggr]_{\eta_i}^{\ee}\nn\\
& &+~{\rm complex~conjugate}.\label{eq:G7}
\end{eqnarray} 
In standard slow roll inflation, the contribution due to $\ei$ vanishes 
with the introduction of the regulator, and it is only the term 
evaluated towards end of inflation that contributes.
Amongst the boundary terms, we have chosen to write this term separately
as it is this contribution that is often taken into account (in slow roll
inflation) through a field 
redefinition~\cite{Maldacena:2002vr,Martin:2011sn,Arroja:2011yj}.
However, as we had mentioned, we do not carry out any field redefinition
and explicitly calculate the contributions due to the bulk as well as the 
boundary terms.

The two terms $G_{8}(\vka,\vkb,\vkc)$ and $G_{9}(\vka,\vkb,\vkc)$ are 
the contributions due to the remaining temporal boundary terms of the 
cubic order action listed in Eq.~(\ref{eq:S3B}).
The contributions $G_{9}(\vka,\vkb,\vkc)$ and $G_{8}(\vka,\vkb,\vkc)$ 
arise due to terms with and without~$\cR'$, respectively. 
They are given by the following expressions:
\begin{subequations}\label{eq:G8-9}
\begin{eqnarray}
G_{8}(\vka,\vkb,\vkc) 
&=& i\,\Mpl^2\, f_{k_1}(\ee)\,f_{k_2}(\ee)\,f_{k_3}(\ee)\,
\l[\f{a}{H}\,f_{k_1}^{\ast}(\eta)\,
f_{k_2}^{\ast}(\eta)\,f_{k_3}^{\ast}(\eta)\r]_{\ei}\nn\\
& & \times\,\Biggl\{54\, (a\,H)^2 +\, 2\,(1-\epsilon_1)\,
(\vka\cdot\vkb +\vka\cdot\vkc + \vkb\cdot\vkc)\nn\\
& &+\, \f{1}{2\,(a\,H)^2}\, \biggl[(\vka\cdot\vkb)\,\skc^2
+ (\vka\cdot\vkc)\,\skb^2+ (\vkb\cdot\vkc)\,\ska^2\biggl]\Biggr\}_{\ei}\nn\\
& & +\,{\rm complex~conjugate},\label{eq:G8}\\
G_{9}(\vka,\vkb,\vkc) 
&=& i\,\Mpl^2\,f_{k_1}(\ee)\,f_{k_2}(\ee)\,f_{k_3}(\ee)\nn \\
& & \times \Bigg\{\f{\epsilon_1}{2\,H^2}\,f_{k_1}^{\ast}(\eta)\,
f_{k_2}^{\ast}(\eta)\,f_{k_3}'^{\ast}(\eta)\,
\biggl[k_1^2 + k_2^2 - \l(\frac{\vka\cdot\vkc}{k_3}\r)^2
-\ \l(\f{\vkb\cdot\vkc}{k_3}\r)^2\biggr]\nn\\
& & -\,\f{a\,\epsilon_1}{H}\,f_{k_1}^{\ast}(\eta)\,
f_{k_2}'^{\ast}(\eta)\,f_{k_3}'^{\ast}(\eta)\,\l[2 - \epsilon_1 
+ \epsilon_1\,\l(\f{\vkb\cdot\vkc}{k_2\,k_3}\r)^2\r]\Bigg\}_{\ei}^{\ee}\nn \\
& & +\,{\rm two~permutations}+ {\rm complex~conjugate}.\label{eq:G9}
\end{eqnarray} 
\end{subequations}
%%%%%%%%%%%%%%%%%%%%%%%%%%%%%%%%%%%%%%%%%%%%%%%%%%%%%%%%%%%%%%%%%%%%%%%%%%%%%%%%
Note that, because $G_{8}(\vka,\vkb,\vkc)$ involves only~$\cR$ (and not 
$\cR'$), its contribution at late times (i.e. at $\ee$) vanishes identically
in any scenario.
Moreover, both the boundary terms $G_{8}(\vka,\vkb,\vkc)$ and
$G_{9}(\vka,\vkb,\vkc)$ generally do not contribute in inflationary 
scenarios that do not have a finite duration. 
But, as we shall see, in the models with kinetically dominated initial 
regimes, these boundary terms can contribute significantly at the initial
time~$\ei$.

The non-Gaussianity parameter~$\fnl(\vka,\vkb,\vkc)$ corresponding to 
the scalar bispectrum $G(\vka,\vkb,\vkc)$ is defined as (see, for 
instance, Refs.~\cite{Martin:2011sn,Hazra:2012yn})
\begin{eqnarray}
\fnl(\vka,\vkb,\vkc)
& =&-\frac{10}{3}\,\frac{1}{\l(2\,\pi\r)^4}\;k_1^3\, k_2^3\, k_3^3\;
G(\vka,\vkb,\vkc)\nn\\
& & \times\, \biggl[k_1^3\,\ps(k_2)\,\ps(k_3) 
+ {\rm two~permutations}\biggr]^{-1},\label{eq:fnl}
\end{eqnarray}
where $\ps(k)$ denotes the scalar power spectrum [cf. Eq.~(\ref{eq:ps})].

%%%%%%%%%%%%%%%%%%%%%%%%%%%%%%%%%%%%%%%%%%%%%%%%%%%%%%%%%%%%%%%%%%%%%%%%%%%%%%%

\subsection{Numerical computation of the scalar bispectrum}

Let us now discuss the numerical evaluation of the scalar bispectrum.
Once the background evolution has been determined, it is a matter of 
arriving at the solution for the modes $f_k$ and then using them to 
compute the integrals~$\cG_{_{C}}(\vka,\vkb,\vkc)$ [cf. Eqs.~(\ref{eq:cG})] 
and the corresponding contributions to the bispectrum.
Evidently, evaluating the contributions due to the boundary terms 
$G_{7}(\vka,\vkb,\vkc)$, $G_{8}(\vka,\vkb,\vkc)$ and $G_{9}(\vka,\vkb,\vkc)$
[cf. Eqs.~(\ref{eq:G7}) and~(\ref{eq:G8-9})] is relatively straightforward 
as it involves no integrals and can be arrived at from the background 
quantities and the modes~$f_k$.

As we had discussed earlier, in the standard slow roll scenario or in 
situations involving brief intermediate departures from slow roll [such
as in the second Starobinsky model (SMII) and punctuated inflation (PI)], 
to arrive at the scalar power spectrum, the modes $f_k$ are evolved 
from the time when $k = 10^2\, \sqrt{z''/z}$ to the time when $k =
10^{-5}\,\sqrt{z''/z}$.
It has been established that it is often adequate to consider the 
evolution of modes over this domain to arrive at the bispectra as 
well (see Refs.~\cite{Chen:2008wn,Hazra:2012yn,Sreenath:2014nca}; in 
this context, also see Refs.~\cite{Sreenath:2013xra,Sreenath:2014nka}).
Since the amplitude of curvature perturbation freezes on super-Hubble 
scales, one finds that the contribution over the domain $k < 10^{-5}\,
\sqrt{z''/z}$ proves to be insignificant.
However, as the bispectrum involves three modes, one has to evolve the 
modes and carry out the integrals from a domain when the smallest of 
the three wave numbers $(k_1,k_2,k_3)$ satisfies the sub-Hubble condition
$k= 10^2\,\sqrt{z''/z}$ until the time when the largest of the three 
satisfy the super-Hubble condition $k = 10^{-5}\,\sqrt{z''/z}$.

In fact, there is yet another point one needs to take into account 
when computing the integrals.
Since the modes oscillate in the sub-Hubble domain, one actually needs 
to introduce a cut-off in order to regulate the integrals involved.
Theoretically, such a cut-off is necessary to identify the correct 
perturbative vacuum (see, for instance, Refs.~\cite{Maldacena:2002vr,
Seery:2005wm}).
Numerically, the cut-off helps us to efficiently compute the integrals.
For an arbitrary triangular configuration of the wave vectors, one often 
works with a democratic cut-off of the form ${\rm exp}\,[-\kappa\,
(k_1+k_2+k_3)/(3\,\sqrt{z''/z})]$, where $\kappa$ is a suitably chosen 
constant.
The value of $\kappa$ is determined by calculating the integrals starting 
from different times inside the Hubble radius and examining the dependence 
of the results for the integrals on the initial time and the value 
of $\kappa$.
It is found that, in most of the cases, if one chooses to integrate from 
$k= 10^2\, \sqrt{z''/z}$, the value of $\kappa\simeq 0.3$ proves to be 
optimal~\cite{Hazra:2012yn,Sreenath:2014nca,Sreenath:2013xra,
Sreenath:2014nka}.
In other words, for $\kappa=0.3$, the values of the integrals prove to be 
independent of how deep inside the Hubble radius the integrals are carried 
out from. 
We use this procedure to calculate the integrals~$\cG_{_{C}}(\vka,\vkb,\vkc)$ 
[cf. Eqs.~(\ref{eq:cG})], the resulting bispectrum $G(\vka,\vkb,\vkc)$ and 
the corresponding non-Gaussianity parameter $\fnl(\vka,\vkb,\vkc)$ in the 
cases of SMII and PI~\cite{Hazra:2012yn}.

\begin{figure}[!t]
\begin{center}
\includegraphics[width=15.0cm]{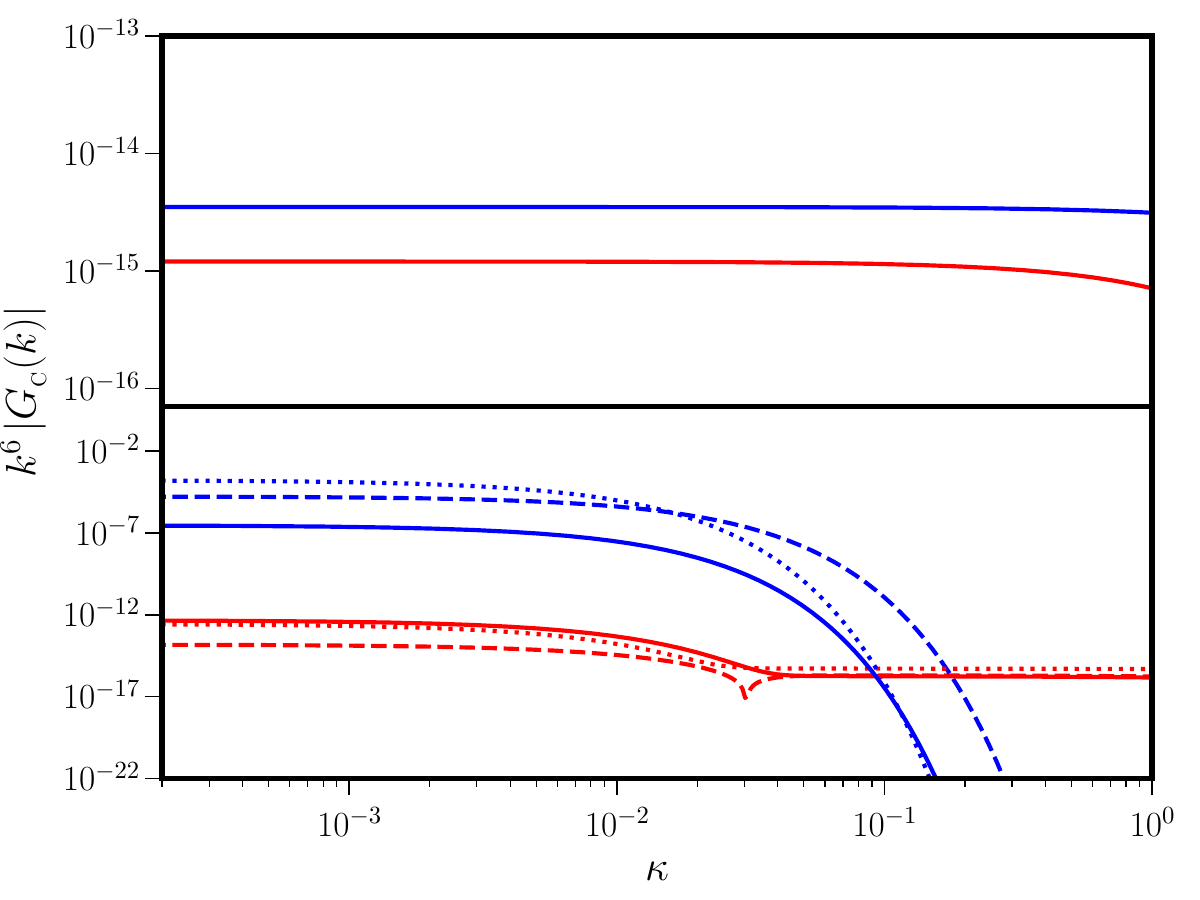}
\end{center}
\vskip -15pt
\caption{The bulk and the boundary contributions to the scalar bispectrum
evaluated numerically in the equilateral limit for the case of the quadratic 
potential with kinetically dominated initial conditions have been plotted as 
functions of the cut-off parameter~$\kappa$.
For highlighting the points we wish to make, we have grouped the six standard 
bulk terms, along with the seventh term, viz. $G_{_{C}}(k)$ with
$C=\l\{1,2,\ldots,7\r\}$ (in red) and the boundary terms, viz. $G_{8}(k)$ 
and $G_{9}(k)$ (in blue). 
We have plotted these quantities for two modes with the wave 
numbers $k=5\times 10^{-5}\,\mpcinv$ (in the top panel) 
and $k=0.1\,\mpcinv$ (in the bottom panel) in the case 
of the model QPa. 
The first of these wave numbers is representative of the modes with 
suppressed power and is always outside the Hubble radius, whereas the 
second corresponds to a typical mode in the nearly scale invariant 
regime that emerges from sufficiently deep inside the sub-Hubble domain
(cf. Fig.~\ref{fig:zeebyz}). 
We have plotted the quantities when the integrals involved have been 
evaluated from $N=0$ (as solid curves) and from the $e$-folds satisfying 
the conditions $k=100\,\sqrt{z''/z}$ and $k=200\,\sqrt{z''/z}$ (as dashed
and dotted curves, respectively), with the latter two being, evidently, 
possible only for the mode with the larger wave number.
Note that, while the quantities are completely insensitive to $\kappa$
for the first mode, the plots suggest the optimal value of the cut-off
parameter to be $\kappa=0.3$ for the second mode.
Also, we should point out that the boundary terms dominate the bulk for 
the mode with the smaller wave number (cf. top panel).  
Moreover, in the case of the mode with the larger wave number, for $\kappa=0.3$, 
the boundary terms cease to be important and the contributions to the bispectrum
are dominated by the bulk terms, as is expected for a mode that emerges from
sufficiently deep inside the Hubble radius.}\label{fig:kappa-kdic}
\end{figure}

But, the scenario with kinetically dominated initial conditions and
variations of it such as its dual and the hard cut-off model pose a 
peculiar problem.
Recall that, in these cases, modes over a certain range of wave numbers are
never inside the Hubble radius (in this context, see Fig.~\ref{fig:zeebyz}).
Therefore, the integrals involving modes over this range do not actually
require a cut-off.
For these modes, we evaluate the integrals from $N=0$ or $N=N_1$ when we 
begin to evolve the perturbations.
When we do so, we find that, the contributions to the scalar bispectrum 
for this range of modes are completely insensitive to the value of the
cut-off parameter~$\kappa$.
This point is illustrated in Fig.~\ref{fig:kappa-kdic} wherein we have 
plotted the contributions to the bispectrum in the equilateral limit (i.e. 
when $k_1=k_2=k_3=k$) due to the bulk and the boundary terms as a function 
of $\kappa$ in the case of QPa.
%%%%%%%%%%%%%%%%%%%%%%%%%%%%%%%%%%%%%%%%%%%%%%%%%%%%%%%%%%%%%%%%%%%%%%%%%%%%%%%
%%%%%%%%%%%%%%%%%%%%%%%%%%%%%%%%%%%%%%%%%%%%%%%%%%%%%%%%%%%%%%%%%%%%%%%%%%%%%%%
Also, in QPa and similar scenarios, for the initial conditions and the best-fit 
values of the parameters we have worked with, we find that  we can impose the 
Bunch-Davies initial condition at $k= 10^2\, \sqrt{z''/z}$ for modes with 
$k \gtrsim 8\times 10^{-3}\, \mpcinv$ (in this context, see Fig.~\ref{fig:zeebyz}). 
As one would have expected, for these modes, the choice of $\kappa=0.3$ turns out 
to be ideal as in the cases of SMII and PI (see Fig.~\ref{fig:kappa-kdic}).
However, since the modes over the range 
$8\times 10^{-5} \lesssim k \lesssim 8\times 10^{-3}\, \mpcinv$
do not spend an adequate amount of time in the sub-Hubble 
domain, we are unable to carry out the exercise described above for 
identifying an apt value of $\kappa$ over this set of wave numbers.
We choose to be democratic, and we work with $\kappa=0.3$ over this range 
of modes as well.
Also, we carry out the integrals from $N=0$ or $N=N_1$ for all the modes 
(viz. for $10^{-5} < k < 1\, {\rm Mpc}^{-1}$) until the time when the largest 
of the three wave numbers involved satisfies the condition $k = 10^{-5}\,
\sqrt{z''/z}$.

\begin{figure}[!t]
\begin{center}
\includegraphics[width=15.0cm]{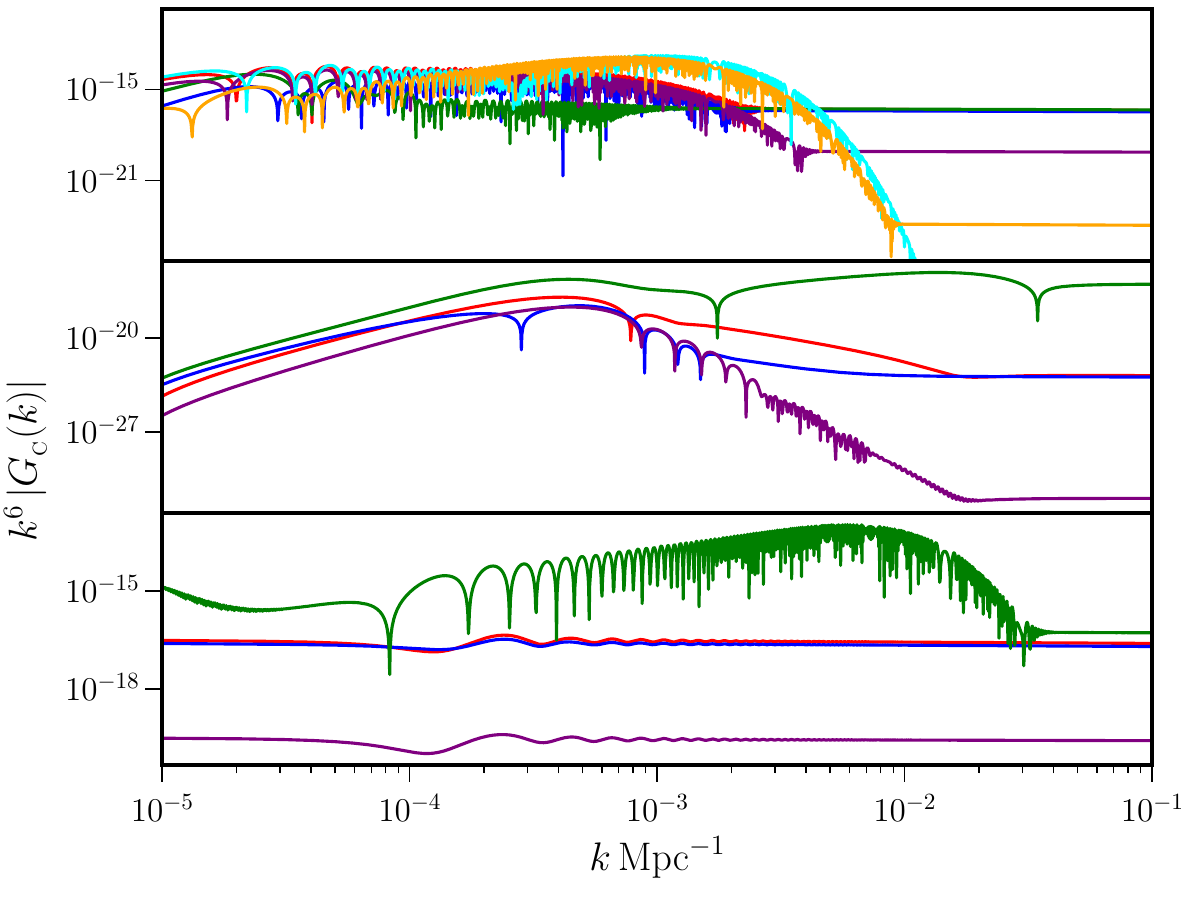}
\end{center}
\vskip -15pt
\caption{The different contributions to the scalar bispectrum in the 
equilateral limit, viz. the bulk terms $G_1(k)+G_3(k)$ (in red), $G_2(k)$ 
(in blue), $G_4(k)+G_7(k)$ (in green), $G_5(k)+G_6(k)$ (in purple) and 
the boundary terms $G_8(k)$ (in cyan) and $G_9(k)$ (in orange), evaluated 
numerically, have been plotted for three models of our interest, 
viz, QPa (on top), PI (in the middle) and SMII (at the bottom).
We should mention that we have made use of the smoothened 
potential~(\ref{eq:sm2}) to evaluate the results numerically 
in the case of SMII. 
Note that, since all the modes of cosmological interest emerge from 
sufficiently inside the Hubble radius in SMII and PI, there arise no 
contributions from the boundary terms in these cases.
However, in the case of QPa, it should be clear that the boundary terms
dominate at small wave numbers.
An interesting point to note regarding PI is that the contribution to
the scalar bispectrum due to $G_4(k)+G_7(k)$ is much higher in magnitude 
when compared to the contribution $G_1(k)+G_2(k)+G_3(k)$ over small scales. 
This is due to the fact that the value of $\epsilon_2$ is relatively larger 
than $\epsilon_1$ during late times in the model. 
Hence, the contributions proportional to $\epsilon_2$ dominate over others 
that are proportional to $\epsilon_1$.
We should also point out the linear growth in $G_4(k)+G_7(k)$ at 
large~$k$ in SMII.
The growth is known to become indefinite in the limit when the quantity 
$\Delta \phi$ in the potential~(\ref{eq:sm2}) vanishes, i.e. when the change 
in the slope of the potential ceases to be smooth and is infinitely abrupt 
as in the original potential~(\ref{eq:v-sm})~\cite{Arroja:2012ae,
Martin:2014kja}.}\label{fig:G}
\end{figure}

In Fig.~\ref{fig:G}, we have plotted the various bulk and boundary 
contributions to the bispectrum for QPa, SMII and PI.
%%%%%%%%%%%%%%%%%%%%%%%%%%%%%%%%%%%%%%%%%%%%%%%%%%%%%%%%%%%%%%%%%%%%%%%%%%%%%%%
%%%%%%%%%%%%%%%%%%%%%%%%%%%%%%%%%%%%%%%%%%%%%%%%%%%%%%%%%%%%%%%%%%%%%%%%%%%%%%%
One finds that, in the equilateral limit, the contributions due to the 
first and the third terms and the contributions due to the fifth and 
the sixth terms have the same form.
Therefore, in the figure, we have plotted the combinations $G_1(k)+G_3(k)$,
$G_2(k)$, $G_4(k)+G_7(k)$\footnote{Note that $G_7(k)$ is not a bulk term
but is actually a boundary term.
Earlier, we had mentioned that the integrals describing the bulk terms do
not contribute when the modes are on super-Hubble scales at late times.
For the term $G_4(k)$, this proves to be true only when the boundary term
$G_7(k)$ is added.
For this reason, often one considers the combination
$G_4(k)+G_7(k)$~\cite{Hazra:2012yn}.}, $G_5(k)+G_6(k)$, $G_8(k)$ and $G_9(k)$.
In the cases of SMII and PI, the boundary terms do not contribute due to
the fact that all the modes of interest emerge from well within the 
Hubble radius.
Also, in these two models, as is well known, it is the contribution due 
to the term $G_4(k)+G_7(k)$ that dominates~\cite{Hazra:2012yn,Sreenath:2014nca}.
This is easy to understand as the term $G_4(k)$ depends on $\epsilon_2'$ 
which grows large for a brief period of time in these scenarios.
In complete contrast, in QPa, one finds that all the contributions to the
bispectrum are roughly of the same order over a wide range of wave numbers.
Moreover, in SMII and PI, all the contributions to the bispectrum are 
enhanced over wave numbers that leave the Hubble radius during the period 
of departure from slow roll inflation.
However, in the case of QPa, the contributions to the scalar bispectrum due 
to the boundary terms dominate the contributions due to the bulk terms over 
a range of large scale modes.
This is a novel result that does not seem to have been noticed earlier
in the literature~\cite{Ragavendra:2019mek}.

%%%%%%%%%%%%%%%%%%%%%%%%%%%%%%%%%%%%%%%%%%%%%%%%%%%%%%%%%%%%%%%%%%%%%%%%%%%%%%%

\subsection{Analytical calculation in the hard cut-off model}

Since it involves only slow roll, the hard cut-off model~(HCO) provides a 
simple situation to evaluate the scalar bispectrum analytically.
In this section, we shall compare the analytical results in this case with 
the corresponding numerical results to highlight the accuracy of our numerical 
computations in situations wherein the initial conditions for a range of modes
are imposed on super-Hubble scales.

It is well known that in slow roll, it is the first, second and the third bulk
terms, viz. $G_{_{C}}(\vk_1,\vk_2,\vk_3)$ with $C=\{1,2,3\}$, that contribute
significantly to the bispectrum.
These bulk terms are characterized by integrals of the form 
[cf. Eqs~.(\ref{eq:cG})]
\begin{subequations}
\begin{eqnarray}
I_1 &=& \int_{\ei}^{\ee} \d\eta\, f_{k_1}(\eta)\,f_{k_2}'(\eta)\,f_{k_3}'(\eta)\,
{\rm e}^{\kappa\,(k_1+k_2+k_3)\,\eta/3} 
+ \, \mbox{\rm two permutations},\\
I_2 &=& \int_{\ei}^{\ee} \d\eta\, f_{k_1}(\eta)\,f_{k_2}(\eta)\,f_{k_3}(\eta)\,
{\rm e}^{\kappa\,(k_1+k_2+k_3)\,\eta/3},
\end{eqnarray}
\end{subequations}
with the modes $f_k$ given by Eq.~(\ref{eq:fk-hco}) in the case of HCO.
Since the initial conditions are imposed on super-Hubble scales for a range of 
modes, apart from these bulk terms, we also need to evaluate the contributions 
due to the boundary terms viz. $G_{_{C}}(\vk_1,\vk_2,\vk_3)$ with $C=\{7,8,9\}$.
While the boundary terms are straightforward to evaluate as they involve no 
integrals, one finds that the above-mentioned integrals are easy to calculate 
as well.

\begin{figure}[!t]
\begin{center}
\hskip -15pt \includegraphics[width=15.0cm]{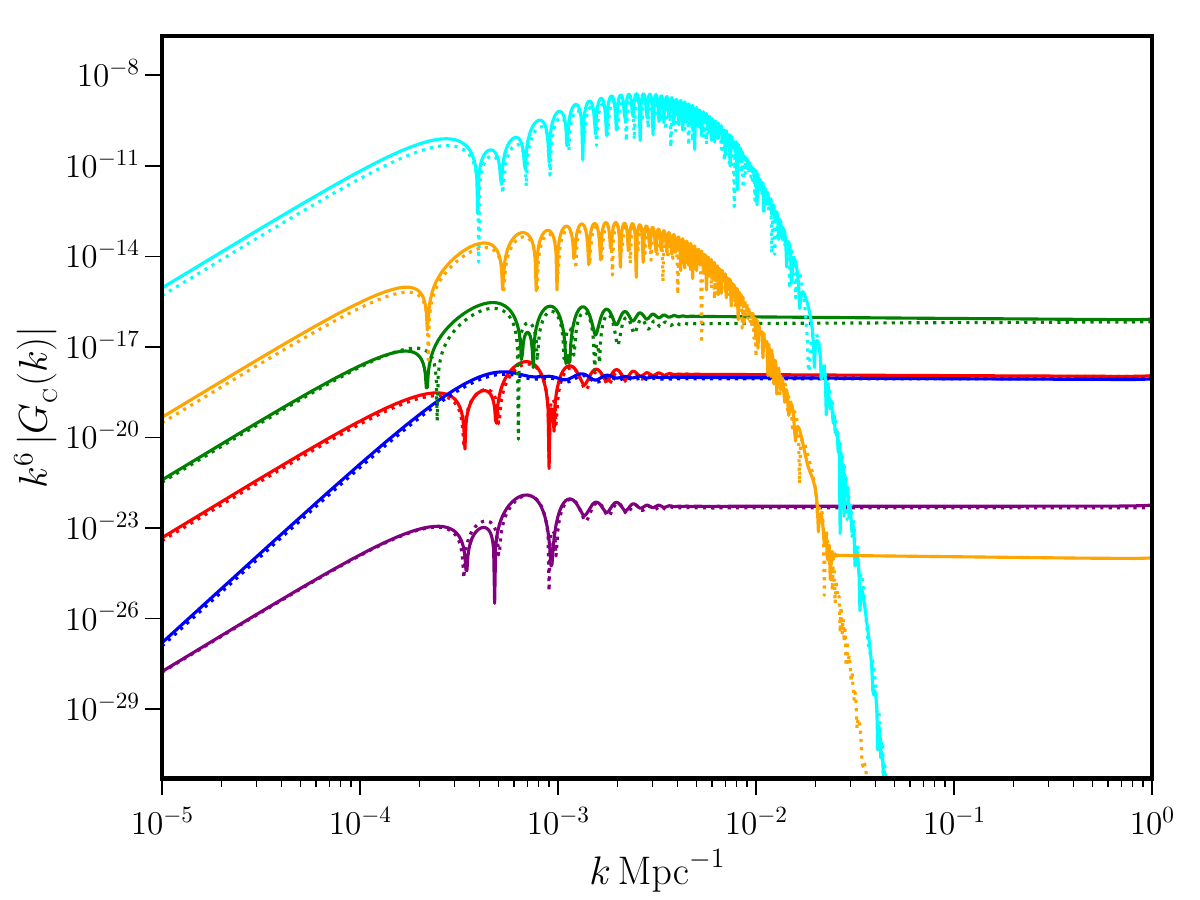}
\end{center}
\vskip -15pt
\caption{The different bulk and boundary contributions to the 
scalar bispectrum, evaluated in the equilateral limit, have been plotted 
for the hard cut-off model (HCO) with the same choices of colors as in the 
previous figure.  
We have plotted the quantities arrived at analytically (as dotted curves) 
as well as numerically (as solid curves).
Clearly, the analytical results match the numerical results quite well.
Moreover, as in the case of QPa plotted in the previous figure, the 
contributions from the boundary terms dominate those due to the bulk 
terms on large scales.}\label{fig:G-hco}
\end{figure}

Note that, in the above integrals, we have introduced the cut-off in a
democratic (in $k_1$, $k_2$, $k_3$) manner that we had discussed earlier.
In Fig.~\ref{fig:G-hco}, we have compared the analytical results for the
different contributions to the bispectrum with the corresponding numerical 
results in the equilateral limit.
To arrive at the numerical results, we have worked with the Starobinsky
potential~(\ref{eq:sm1}) and have started the evolution on the inflationary 
attractor, as we had described in Subsec.~\ref{sec:hco} wherein we had 
discussed the scalar power spectrum arising in the model.
%%%%%%%%%%%%%%%%%%%%%%%%%%%%%%%%%%%%%%%%%%%%%%%%%%%%%%%%%%%%%%%%%%%%%%%%%%%%%%%
%%%%%%%%%%%%%%%%%%%%%%%%%%%%%%%%%%%%%%%%%%%%%%%%%%%%%%%%%%%%%%%%%%%%%%%%%%%%%%%
It is clear that the analytical results match well with the numerical
results indicating the extent of accuracy of the numerical procedures 
we have adopted.
As in the cases of QP and SMI, we find that the boundary terms, in 
particular~$G_8(k)$, dominate at suitably small wave numbers.

%%%%%%%%%%%%%%%%%%%%%%%%%%%%%%%%%%%%%%%%%%%%%%%%%%%%%%%%%%%%%%%%%%%%%%%%%%%%%%%%

\section{Amplitude and shape of the non-Gaussianity parameter}\label{sec:fnl}

Having obtained the scalar bispectrum, let us now turn to understand the 
amplitude and shape of the corresponding non-Gaussianity parameter~$\fnl$.
In the next section, we shall discuss the behavior of the parameter in the
so-called squeezed limit wherein it is expected to be expressed completely 
in terms of the scalar spectral index.
In this section, we shall discuss the behavior in the equilateral limit as
well as the complete shape, which is often illustrated in the form of
density plots.

\par

\begin{figure}[!t]
\begin{center}
\includegraphics[width=8.50cm]{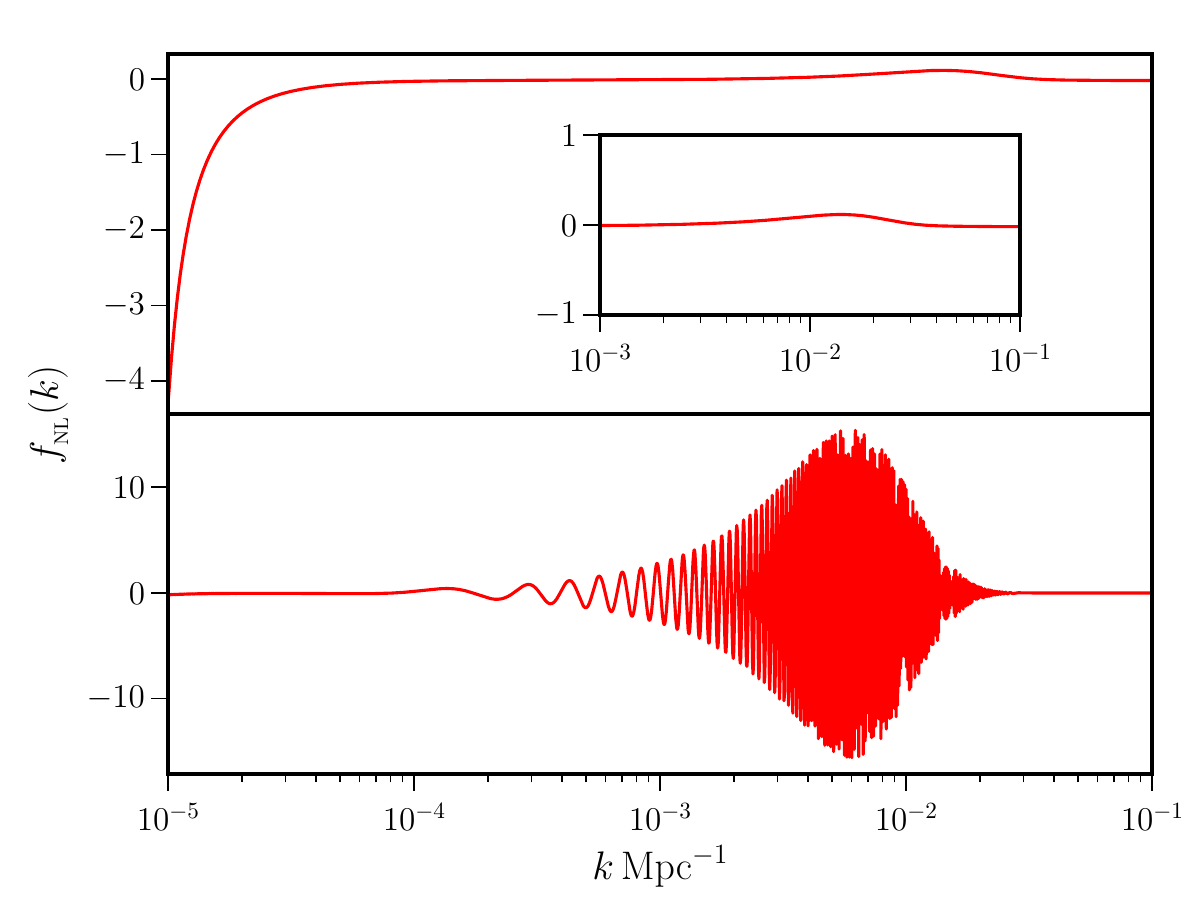}
\includegraphics[width=8.50cm]{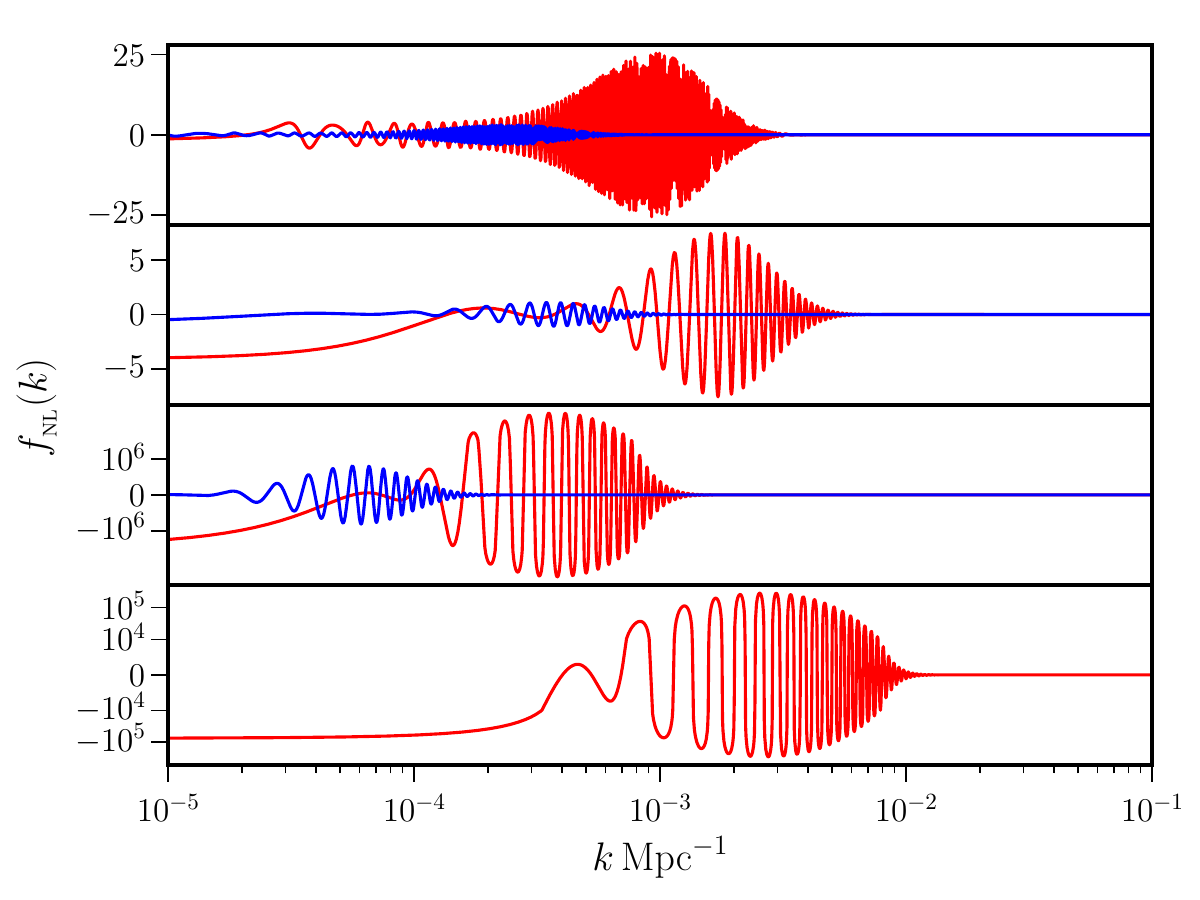}
\end{center}
\vskip -15pt
\caption{The scalar non-Gaussianity parameter~$\fnl$ computed in the 
equilateral limit has been plotted for all the models of our interest: PI and 
SMII (in the top and bottom panels on the left), QPa, QPb and QPc (in the top 
three panels on the right, respectively, as red curves), SMIa, SMIb 
and SMIc (in the top three panels on the right, in blue) 
and, lastly, HCO (in the bottom panel on the right).
Note that the scalar power spectrum appears in the denominator in the 
definition of $\fnl$ [cf. Eq.~(\ref{eq:fnl})].
The spike in the amplitude of $\fnl$ in the case of PI 
(at the left extreme of the figure) arises due to the sharp 
drop in the power spectrum (in this context, see Fig.~\ref{fig:ps}). 
The oscillations with increasing amplitude at larger wave numbers in the case 
of SMII is caused due to the contribution from $G_4(k)+G_7(k)$, which rises
linearly before eventually dying down (cf. Fig.~\ref{fig:G}).
Such a behavior occurs due to the sharp transition that occurs in the evolution 
of the field as it crosses the discontinuity in the derivative of the potential.
Also note that the maximum amplitude of $\fnl$ is larger in QPa and SMIa 
when compared to QPb and SMIb (also see Ref.~\cite{Ragavendra:2019mek}).
This can be partly attributed to the larger initial velocity of the background 
scalar field when the initial conditions are imposed on the perturbations.
Moreover, interestingly, we find that the amplitude of $\fnl$ is larger in 
the case of QP than SMI.
Lastly, the amplitude of $\fnl$ in the cases of QPc, SMIc and HCO are extremely
large, indicating that these models are unlikely to be viable in the light of 
the constraints on $\fnl$ from Planck.}\label{fig:fnl-el}
\end{figure}

Let us first consider the equilateral limit.
In Fig.~\ref{fig:fnl-el}, we have illustrated the behavior of the parameter 
$\fnl$ in the equilateral limit in the different models of our interest.
%%%%%%%%%%%%%%%%%%%%%%%%%%%%%%%%%%%%%%%%%%%%%%%%%%%%%%%%%%%%%%%%%%%%%%%%%%%%%%%
%%%%%%%%%%%%%%%%%%%%%%%%%%%%%%%%%%%%%%%%%%%%%%%%%%%%%%%%%%%%%%%%%%%%%%%%%%%%%%%%
Recall that, according to the most recent constraints from Planck:
$\fnl^{\rm local}= -0.9 \pm 5.1$, $\fnl^{\rm equil}= -26 \pm 47$ 
and $\fnl^{\rm ortho}= -38 \pm 24$~\cite{Akrami:2019izv}.
Among the models we have considered, we find that the parameter $\fnl$ 
is very large in the cases of QPc, SMIc and HCO.
In fact, these scenarios are likely to be inconsistent with the most recent 
constraints on the parameter.
The models SMII and PI also lead to relatively large value of $\fnl$, but
this can attributed to the sharp drop in the scalar power spectra over the 
relevant scales rather than a rise in the amplitude of the bispectrum. 
As we shall discuss in the concluding section, it seems urgent to arrive
at a template for the bispectrum in models such as PI in order to be able
to compare it with the CMB data at the level of three-point functions.

\par

\begin{figure}[!t]
\begin{center}
\includegraphics[width=7.50cm]{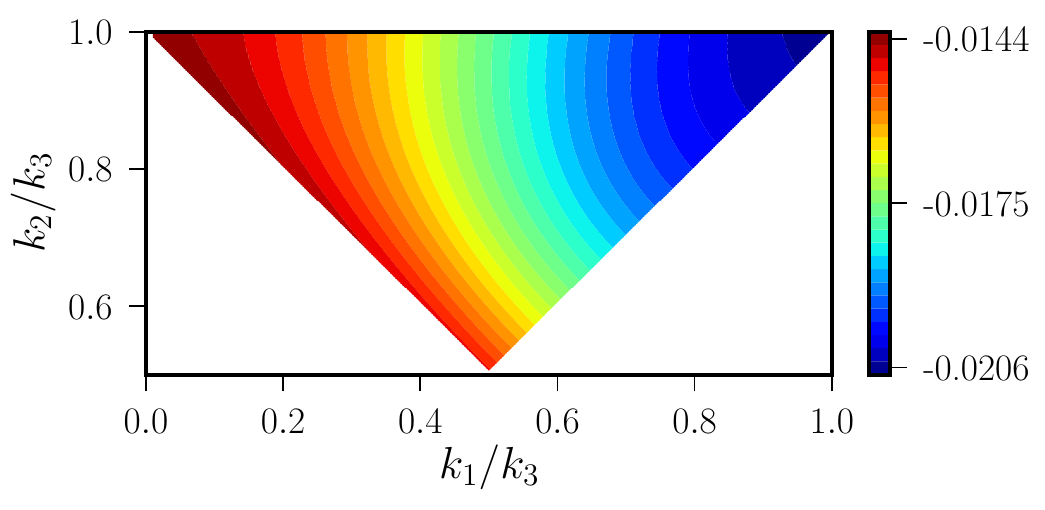}
\includegraphics[width=7.50cm]{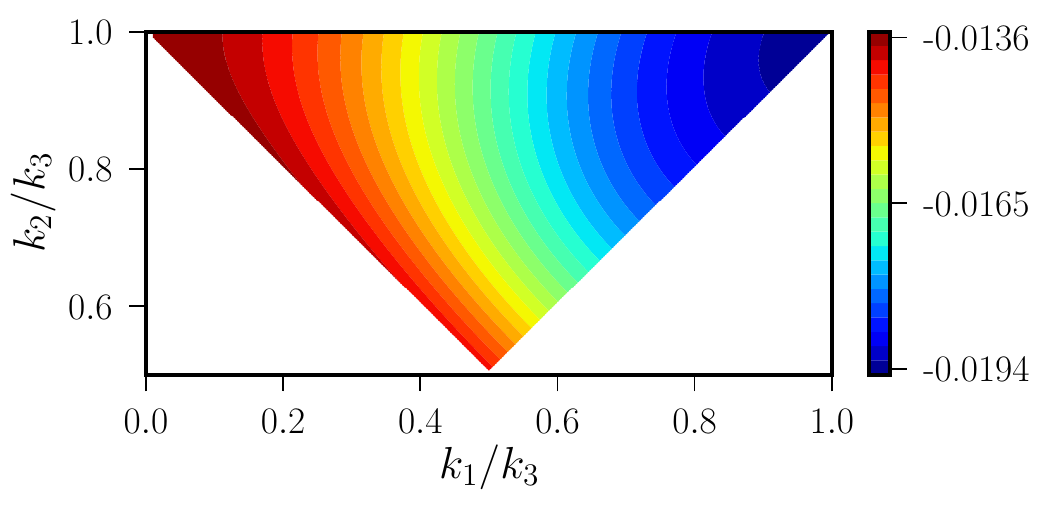}\\
\includegraphics[width=7.50cm]{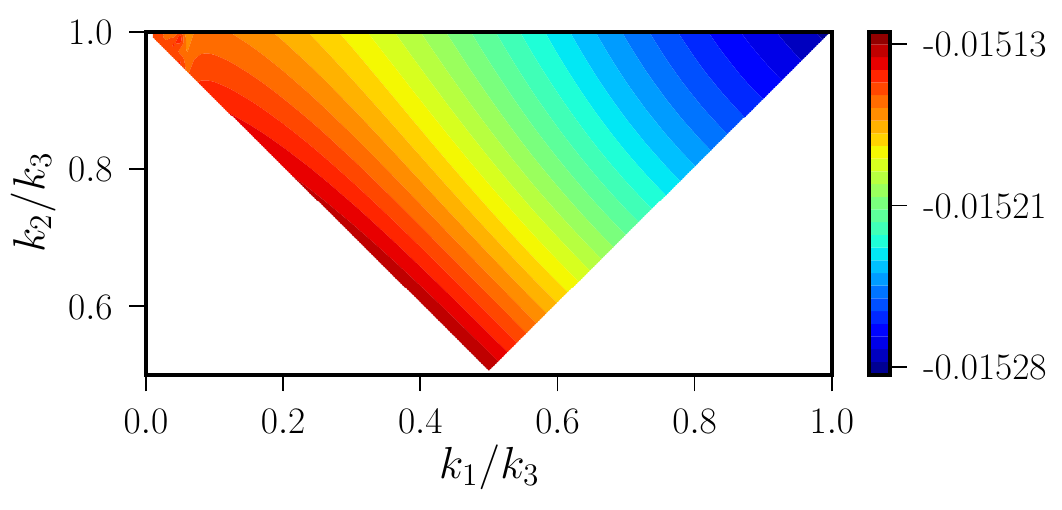}
\includegraphics[width=7.50cm]{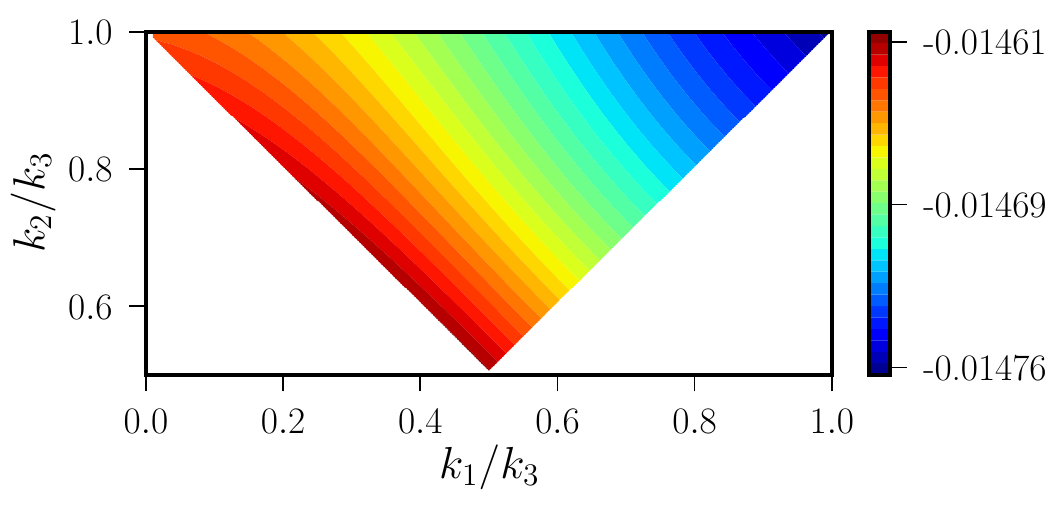}\\
\includegraphics[width=7.50cm]{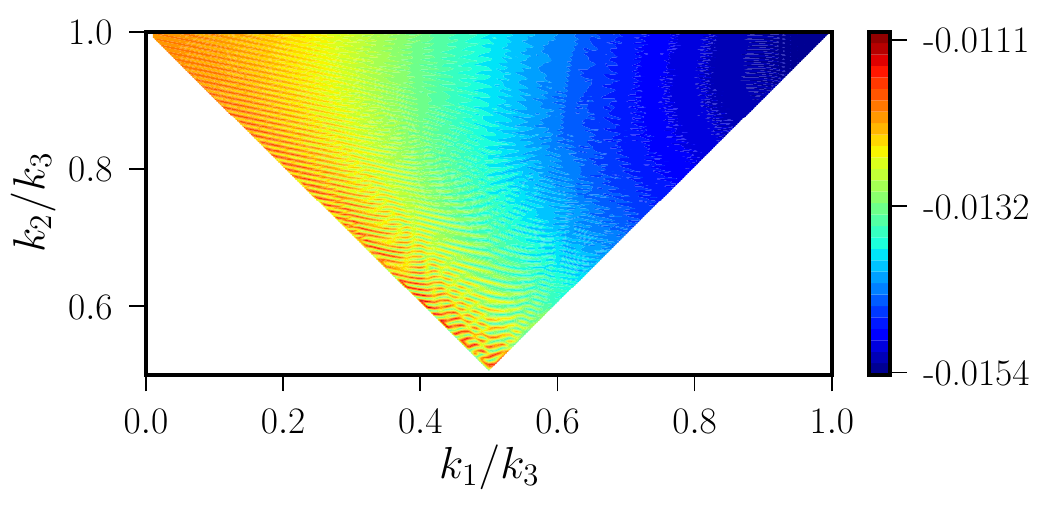}
\includegraphics[width=7.50cm]{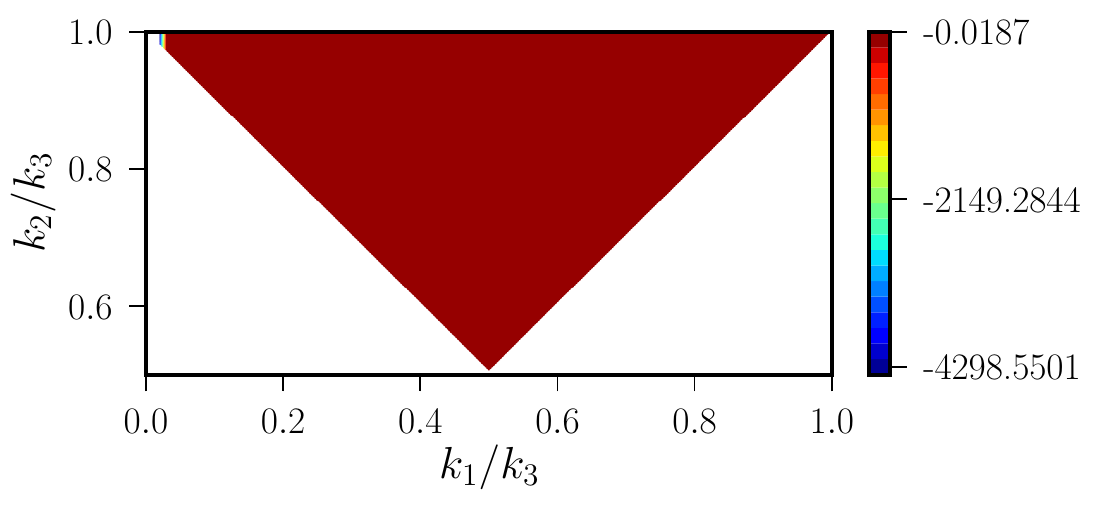}
\end{center}
\vskip -15pt
\caption{The amplitude and shape of the non-Gaussianity 
parameter~$\fnl(\vka,\vkb,\vkc)$ has been illustrated
as density plots for the various models of our interest (QPa, SMIa and 
SMII from top to bottom on the left, and QPb, SMIb and PI in the same 
order on the right) as a function of $k_1/k_3$ and $k_2/k_3$.
We have chosen $k_3$ to be the pivot scale in all the plots.
Note that the top right and left corners of the triangles correspond 
to the non-Gaussianity parameter in the equilateral and squeezed limits, 
evaluated at the pivot scale, respectively.}
\label{fig:fnl}
\end{figure}

In Fig.~\ref{fig:fnl}, we have illustrated the complete shape of the 
scalar non-Gaussianity parameter $\fnl(\vka,\vkb,\vkc)$ that arises in
the various models of our interest in the form of density plots,
as is usually done.
%%%%%%%%%%%%%%%%%%%%%%%%%%%%%%%%%%%%%%%%%%%%%%%%%%%%%%%%%%%%%%%%%%%%%%%%%%%%%%%%
%%%%%%%%%%%%%%%%%%%%%%%%%%%%%%%%%%%%%%%%%%%%%%%%%%%%%%%%%%%%%%%%%%%%%%%%%%%%%%%%
We should mention that the density plots of~$\fnl$ have been 
computed with $k_3$ set to be the pivot scale 
(recall that, $k_\ast=0.05\, \mathrm{Mpc}^{-1}$).
For the models QPa and QPb, we find that the non-Gaussianity parameter around
the pivot scale is equilateral in shape corresponding to the slow roll value 
of $\fnl \simeq 2\times10^{-2}$ in the equilateral limit 
(i.e. when $k_1=k_2 =k_3$, corresponding to the top right corner of the triangular 
density plots). These results should be compared with the corresponding values 
in Fig.~\ref{fig:fnl-el} around the pivot scale. 
The suppression in the scalar power spectrum, which occurs at roughly two decades 
away in wave number, does not affect the shape of $\fnl$ around the pivot scale. 
In the cases of SMIa and SMIb, we see a roughly similar behavior with a slightly 
lesser amplitude of $\fnl$ as expected in SMI models.
In the case of SMII, we see small but persistent oscillations, with $\fnl 
\simeq 10^{-2}$, throughout the range of wave numbers around pivot scale.
This can be understood from the behavior of the contribution $G_4(k)+G_7(k)$ 
in the model. 
In PI, the bispectrum is largely local in shape with a sharp increase in amplitude 
occurring at wave numbers around the location where the scalar spectrum 
exhibits a sharp drop in power.

%%%%%%%%%%%%%%%%%%%%%%%%%%%%%%%%%%%%%%%%%%%%%%%%%%%%%%%%%%%%%%%%%%%%%%%%%%%%%%%%

\section{Validity of the consistency relation}\label{sec:cr}

Let us now turn to the behavior of the three-point functions in the squeezed 
limit wherein one of the three wave numbers is much smaller than the other
two~\cite{Maldacena:2002vr,Creminelli:2004yq,Cheung:2007sv,Sreenath:2014nca}.
Since the amplitude of the long wavelength mode freezes on super-Hubble 
scales during inflation, it can be treated as part of the background.
Consequently, one finds that, in such a limit, the three-point functions 
generated during inflation can be expressed entirely in terms of the 
two-point functions through the so-called consistency relation.
In the squeezed limit, the scalar bispectrum is expected to reduce to 
the following form (see, for instance, Ref.~\cite{Sreenath:2014nca}):
\begin{equation}
\lim_{k_1 \to 0}G(\vka,\vk,-\vk)
= -\f{(2\,\pi)^{4}}{4\, k_1^3\,k^3}\, 
\l[\ns(k) - 1\r]\, \ps(k_1)\,\ps(k),\label{eq:cr-rrr}
\end{equation}
where $\ns(k)=1+\l[\d\,{\rm ln}\,\ps(k)/\d\, {\rm ln}\, k\r]$ is the scalar
spectral index, and it should be clear that we have considered ${\bm \vka}$ 
to be the squeezed mode.
Upon substituting the above expression in the definition~(\ref{eq:fnl}) 
for the non-Gaussianity parameter $\fnl(\vka,\vkb,\vkc)$, we find that 
we can express the consistency relation in the squeezed limit as 
follows:
\begin{equation}
\lim_{k_1 \to 0}\,\fnl(\vka,\vk,-\vk) 
= \frac{5}{12}\,\l[\ns(k)-1\r]
\equiv \fnl^{_{\rm CR}}(k).\label{eq:cr}
\end{equation}

\begin{figure}[!t]
\begin{center}
\includegraphics[width=8.50cm]{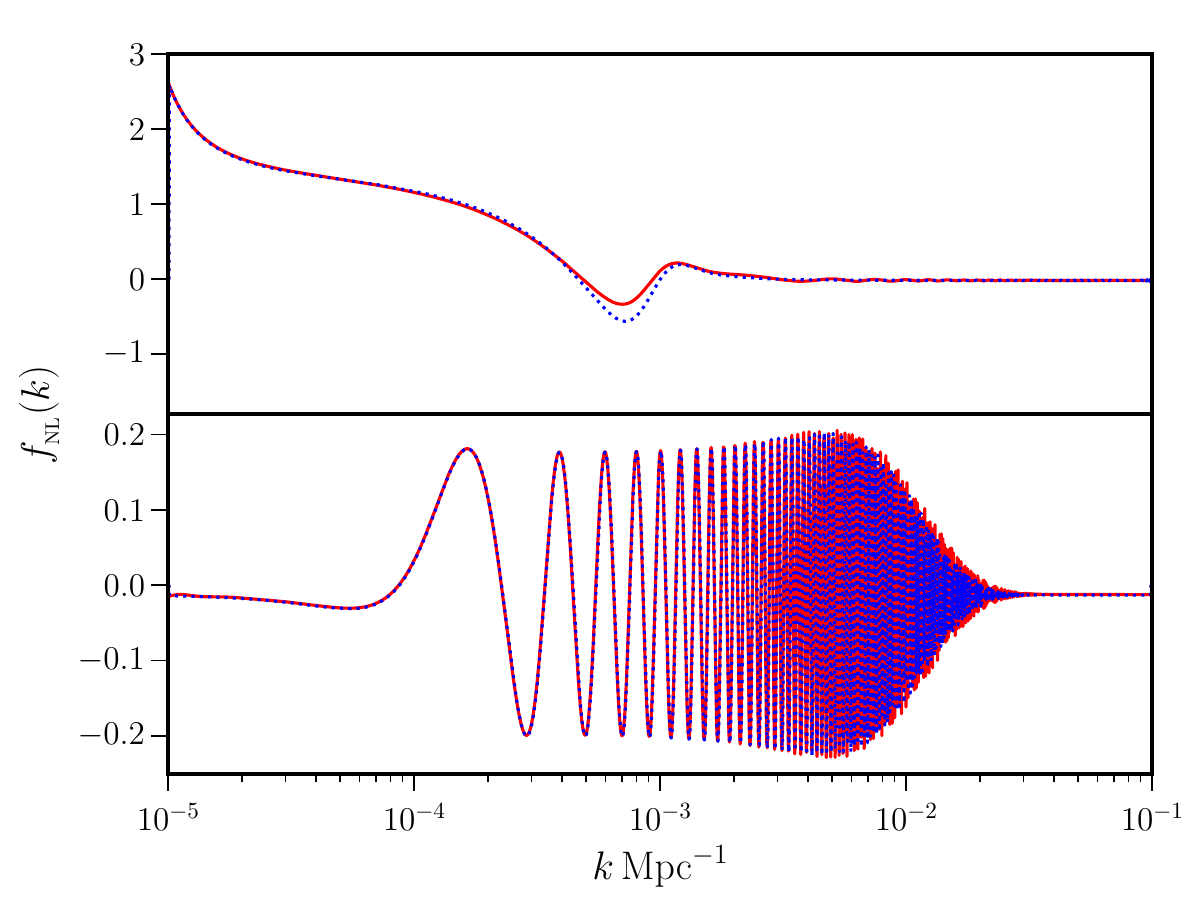}
\includegraphics[width=8.50cm]{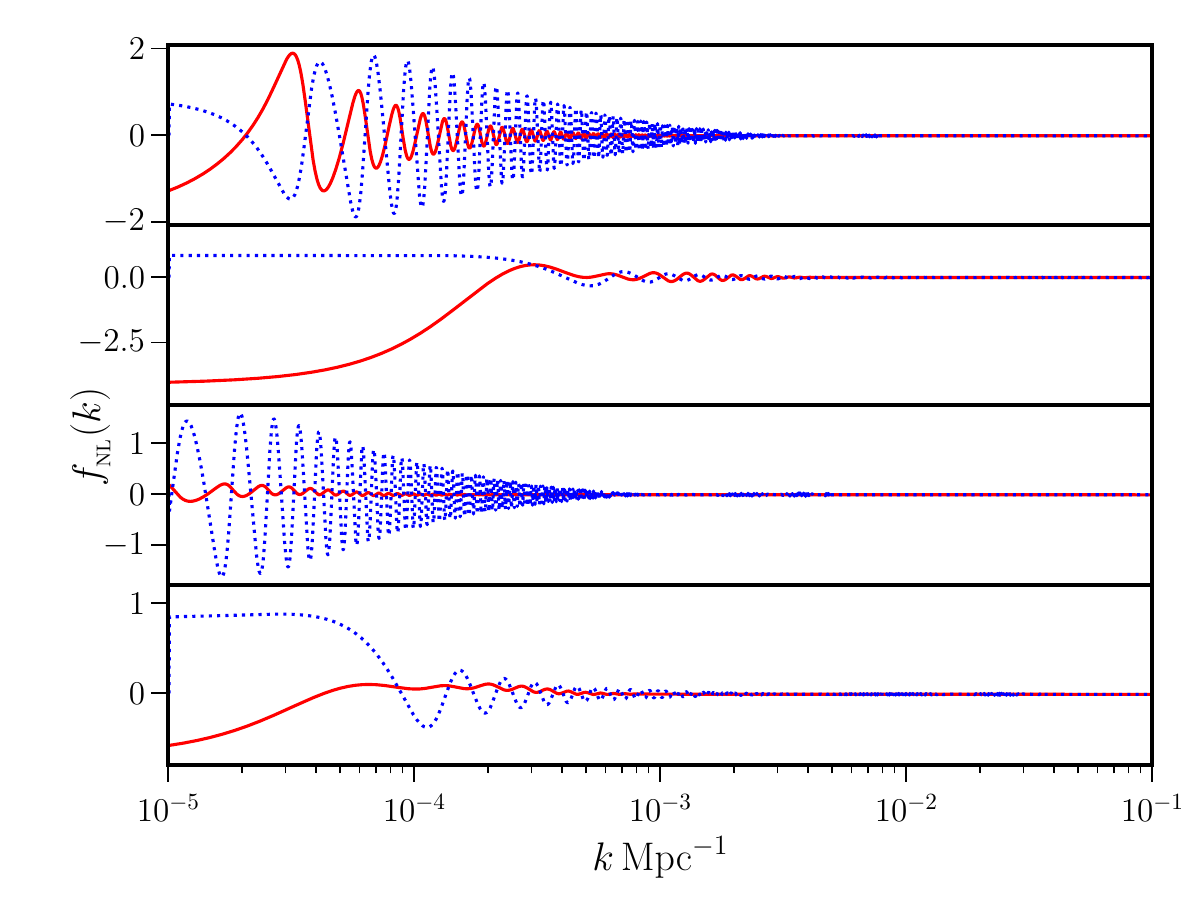}
\end{center}
\vskip -15pt
\caption{The non-Gaussianity parameter $\fnl(k)$ in the squeezed limit has been
plotted (in red) for the cases of PI, SMII (top and bottom panels, on the left), 
QPa, QPb, SMIa and SMIb (panels from top to bottom in that order, on the right).
We have also plotted the quantity $\fnl^{_{\rm CR}}(k)$ [cf. Eq.~(\ref{eq:cr})],
determined completely by the scalar spectral index, for each of these models (as 
dotted blue curves).
Clearly, the consistency condition~(\ref{eq:cr}) is satisfied in PI and SMII (as
is evident from the figure on the left) even over wave numbers wherein there arise
strong departures from near scale invariance in the power and bi-spectra.
In complete contrast, in QPa, QPb, SMIa and SMIb, the consistency condition is 
violated at large scales (as should be clear from the figure on the right), but 
it is eventually restored at the small scales (in this context, also see our 
earlier work~\cite{Ragavendra:2019mek}).
We find that the behavior of $\fnl$ is similar in the cases of QPc, SMIc and 
HCO. Hence, we have not plotted them here. 
The difference arises only in the magnitude of $\fnl$ over large scales 
where the consistency condition is violated.}\label{fig:fnl-cr}
\end{figure}

With the results we have obtained, it is straightforward to examine if 
the consistency relation is satisfied in the models of our interest.
Actually, it has already been established that the consistency relation 
is satisfied in SMII and PI despite the strong departures from slow roll,
as reflected in the sharp features in the power spectra and bispectra 
(see Fig.~\ref{fig:fnl-cr}; in this context, also see 
Refs.~\cite{Martin:2011sn,Sreenath:2014nca}).
%%%%%%%%%%%%%%%%%%%%%%%%%%%%%%%%%%%%%%%%%%%%%%%%%%%%%%%%%%%%%%%%%%%%%%%%%%%%%%%%
%%%%%%%%%%%%%%%%%%%%%%%%%%%%%%%%%%%%%%%%%%%%%%%%%%%%%%%%%%%%%%%%%%%%%%%%%%%%%%%%
However, in the case of the scenarios with kinetically dominated initial 
conditions, we find that the consistency condition is violated on
large scales where the scalar power spectrum exhibits a suppression. 
This should be clear from Fig.~\ref{fig:fnl-cr} wherein we have 
plotted the non-Gaussianity parameter $\fnl(k)$ in the squeezed limit 
as well as the quantity~$\fnl^{_{\rm CR}}(k)$ [cf. Eq.~(\ref{eq:cr})]
for most of the models we have been interested in.
We find that the consistency relation begins to be satisfied in these cases 
only at small scales (for $k\gtrsim 8\times 10^{-3}\,{\rm Mpc}^{-1}$) which 
emerge from sufficiently deep inside the Hubble radius [say, from $k\simeq 
10^{2}\,\sqrt{\vert z''/z}\vert$] after slow roll inflation has set in.
Evidently, the violation of the consistency condition is associated with the
fact that the Bunch-Davies initial condition on the large scale modes are
imposed when they are outside the Hubble radius.
We should mention here that the violation of the consistency condition at 
large scales that we encounter is partly similar to the violation of the 
condition noticed earlier in the case of non-attractor 
inflation~\cite{Namjoo:2012aa,Martin:2012pe,Chen:2013aj,Cai:2016ngx,Cai:2017bxr}.

%%%%%%%%%%%%%%%%%%%%%%%%%%%%%%%%%%%%%%%%%%%%%%%%%%%%%%%%%%%%%%%%%%%%%%%%%%%%%%%%

\section{Summary and scope}\label{sec:dis}

At the level of the power spectrum, all the models we have considered here,
viz. models with kinetically dominated initial conditions, their dual, the 
hard cut-off model, the second Starobinsky model and punctuated inflation, 
lead to a suppression of power on large scales. 
Naively, one would have expected that non-Gaussianities would help us 
discriminate between the different models, and we find that indeed 
they do.
Though there arise some differences in the overall amplitude of the scalar 
bispectra in the various models, the crucial distinction seems to be their
behavior in the squeezed limit.
While the consistency condition is satisfied in PI and SMII over all the 
modes of cosmological interest, in the models with initial kinetic domination,
their dual and HCO, the consistency relation is found to be violated on 
large scales for the modes that always remain in the super-Hubble regime.
However, as in the cases of PI and SMII, in QP, SMI and HCO, the consistency 
relation is satisfied for the small scales modes which evolve from the 
sub-Hubble regime.

Models such as punctuated inflation or the second Starobinsky model may 
be considered to be more appealing theoretically than the models with 
kinetically dominated initial conditions.
However, the data can help us evaluate the performance of the models and 
rule in favor of one over the other.
In order to compare with the CMB data at the level of the bispectrum, it will
be useful to obtain an analytical template for the scalar bispectrum
(in this context, see, for example, Refs.~\cite{Adshead:2011bw,Basu:2019jew}).
While there have been efforts to reproduce the power spectra analytically in the 
case of models with kinetically dominated initial conditions (in this context, 
see Ref.~\cite{Contaldi:2003zv}), these analytical calculations seem to 
underestimate the amplitude of the oscillations that arise as the spectrum turns 
scale invariant.
In the context of PI, there seems to have been no effort at all to arrive at 
the power spectrum analytically.
We are currently working on evaluating the spectra as well as the bispectra
analytically in PI as well as in models with kinetically dominated initial 
conditions with the aim of eventually comparing these models with the CMB
data at the level of bispectra~\cite{Sohn:2019rlq}.

%%%%%%%%%%%%%%%%%%%%%%%%%%%%%%%%%%%%%%%%%%%%%%%%%%%%%%%%%%%%%%%%%%%%%%%%%%%%%%%%

\section*{Acknowledgements}

The authors wish to thank Xingang Chen, Dhiraj Hazra and Takahiro Tanaka 
for discussions and comments on the manuscript. 
HVR would like to thank the Indian Institute of Technology Madras (IIT-M), 
Chennai, India, for financial support through half-time research assistantship.
DC would like to thank the Tata Institute of Fundamental Research (TIFR), 
Mumbai, India, for financial support. 
DC's work is also supported by STFC grant ST/T000813/1.
The authors wish to acknowledge the use of the High Performance Computing 
Environment at IIT-M and the cluster computing facilities at TIFR.
LS also wishes to acknowledge support from the Science and Engineering Research  
Board, Department of Science and Technology, Government of India, through the 
Core Research Grant CRG/2018/002200.

%%%%%%%%%%%%%%%%%%%%%%%%%%%%%%%%%%%%%%%%%%%%%%%%%%%%%%%%%%%%%%%%%%%%%%%%%%%%%%%%

\appendix

%%%%%%%%%%%%%%%%%%%%%%%%%%%%%%%%%%%%%%%%%%%%%%%%%%%%%%%%%%%%%%%%%%%%%%%%%%%%%%%%

\section{Signatures of initial kinetic domination across models}\label{app:ii}

\begin{figure}[!t]
\begin{center}
\includegraphics[width=15.0cm]{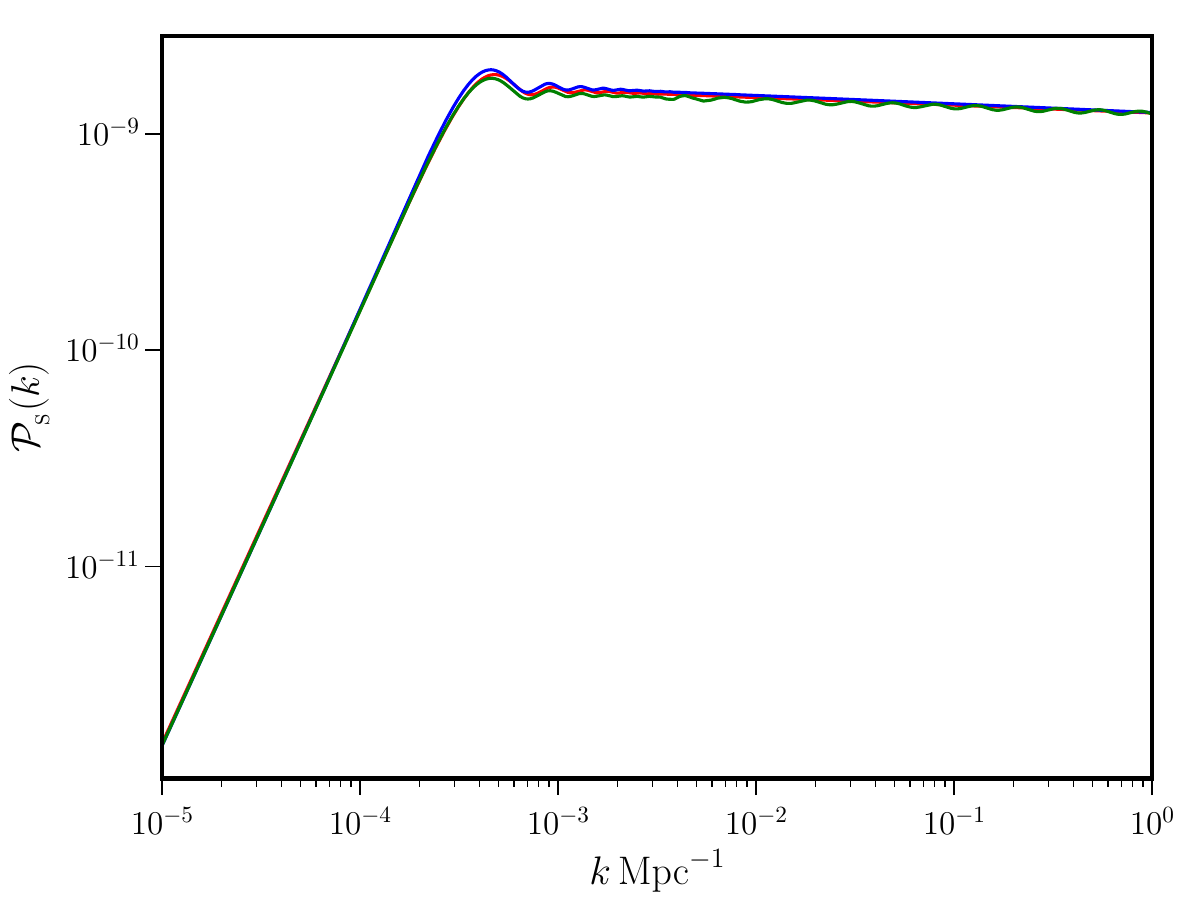}
\end{center}
\vskip -15pt
\caption{The scalar power spectra in a small field inflationary model (in blue) 
and the axion monodromy model (in green) with kinetically dominated initial 
conditions have been plotted along with the power spectrum in the case of QPa
(in red).
The parameters have been chosen so that the features of the power spectra match
at large scales. We should clarify we have not carried out a comparison of the
small field model and the axion monodromy model against the CMB
data.}\label{fig:kdi-vs-axm-vs-sfi}
\end{figure}

\begin{figure}[!t]
\begin{center}
\includegraphics[width=15.0cm]{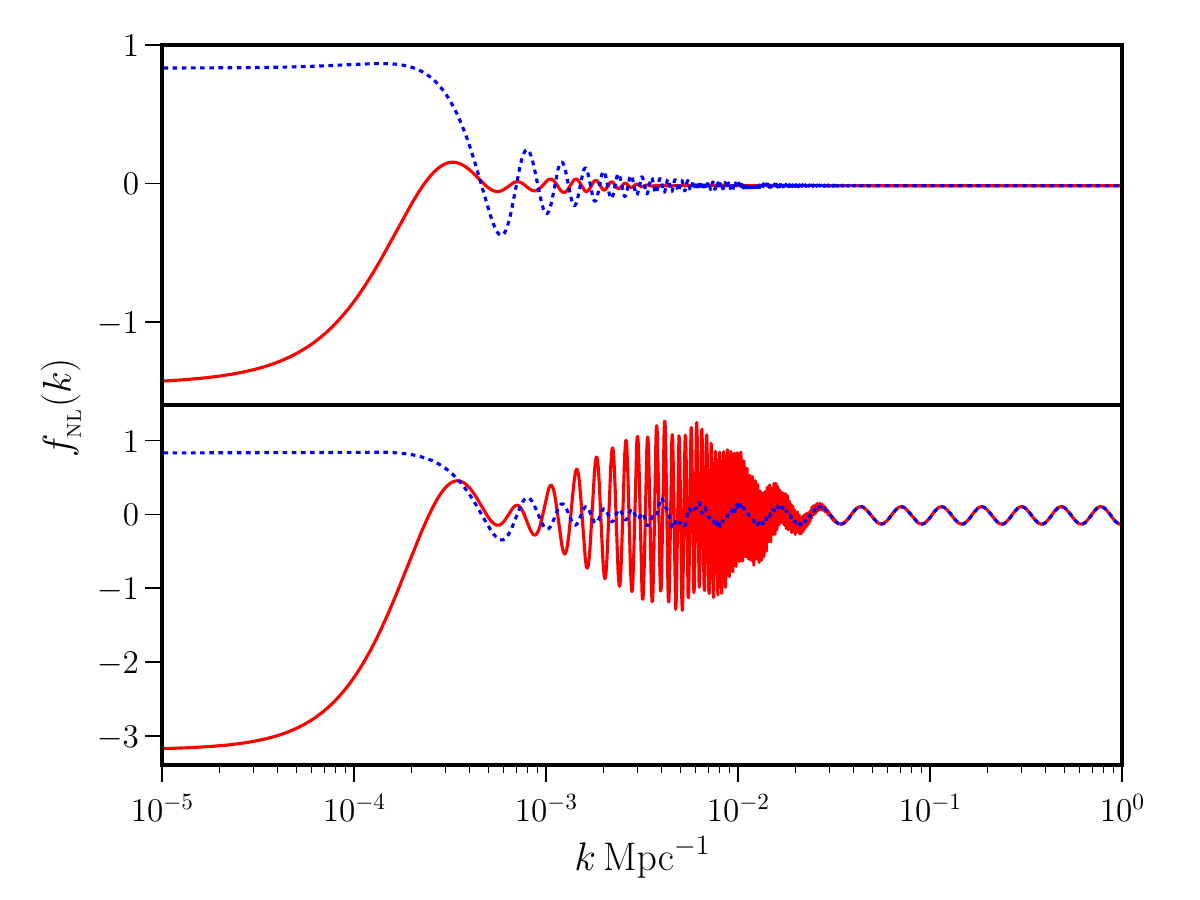}
\end{center}
\vskip -15pt
\caption{The behavior of the scalar non-Gaussianity parameter $\fnl$ in the 
squeezed limit has been plotted (in red) for the small field inflationary 
model (on top) and the axion monodromy model (at the bottom).
Just as we had done earlier, we have also plotted the quantity $\fnl^{_{\rm CR}}$
(in blue).
As in the cases of QP and SMI, while the consistency condition is violated at
large scales, it is restored at small scales.
This is clearly evident in the case of the axion monodromy model which is known 
to exhibit oscillations in the power spectrum as well as bispectrum even at 
small scales.}\label{fig:fnl-sfi-axm}
\end{figure}

To illustrate that the imprints of initial kinetic domination arise across all
inflationary modes, in this appendix, we shall consider two other models of
inflation, viz. a small field model and so-called the axion monodromy model, 
which are described by the following potentials:
\begin{subequations}
\begin{eqnarray}
V(\phi) &=& V_0\,\l[1-\l(\f{\phi}{\phi_0}\r)^4\r],\\
V(\phi) &=& \mu^3\,\l[\phi + b\,\phi_0\,
{\rm cos}\,\l(\frac{\phi}{\phi_0}\r)\r].
\end{eqnarray}
\end{subequations}
We work with parameters and initial conditions for the background such that the
power spectra are COBE normalized around the pivot scale and the suppression on
large scales occurs as in QPa.
The corresponding scalar power spectra are illustrated in Fig.~\ref{fig:kdi-vs-axm-vs-sfi},
and it is clear that, despite the different choice of potentials, the power spectra
have the same shape at large and small scales across models.
%%%%%%%%%%%%%%%%%%%%%%%%%%%%%%%%%%%%%%%%%%%%%%%%%%%%%%%%%%%%%%%%%%%%%%%%%%%%%%%%
%%%%%%%%%%%%%%%%%%%%%%%%%%%%%%%%%%%%%%%%%%%%%%%%%%%%%%%%%%%%%%%%%%%%%%%%%%%%%%%%
In Fig.~\ref{fig:fnl-sfi-axm}, we have plotted the behavior of the non-Gaussianity
parameter $\fnl$ in the squeezed limit in these cases.
%%%%%%%%%%%%%%%%%%%%%%%%%%%%%%%%%%%%%%%%%%%%%%%%%%%%%%%%%%%%%%%%%%%%%%%%%%%%%%%%
%%%%%%%%%%%%%%%%%%%%%%%%%%%%%%%%%%%%%%%%%%%%%%%%%%%%%%%%%%%%%%%%%%%%%%%%%%%%%%%%
Clearly, the behavior of the parameter is similar to that encountered in the 
cases of QP and SMI. 
The restoration of the consistency condition is well illustrated in the case 
of the axion monodromy model, wherein both the power and bispectra exhibit 
continued oscillations even at small scales~\cite{Aich:2011qv,Sreenath:2014nca}.

%%%%%%%%%%%%%%%%%%%%%%%%%%%%%%%%%%%%%%%%%%%%%%%%%%%%%%%%%%%%%%%%%%%%%%%%%%%%%%%%

\section{Constraints on the cosmological parameters}\label{app:ccp}

\begin{figure*}
\begin{center}
\includegraphics[width=8.50cm]{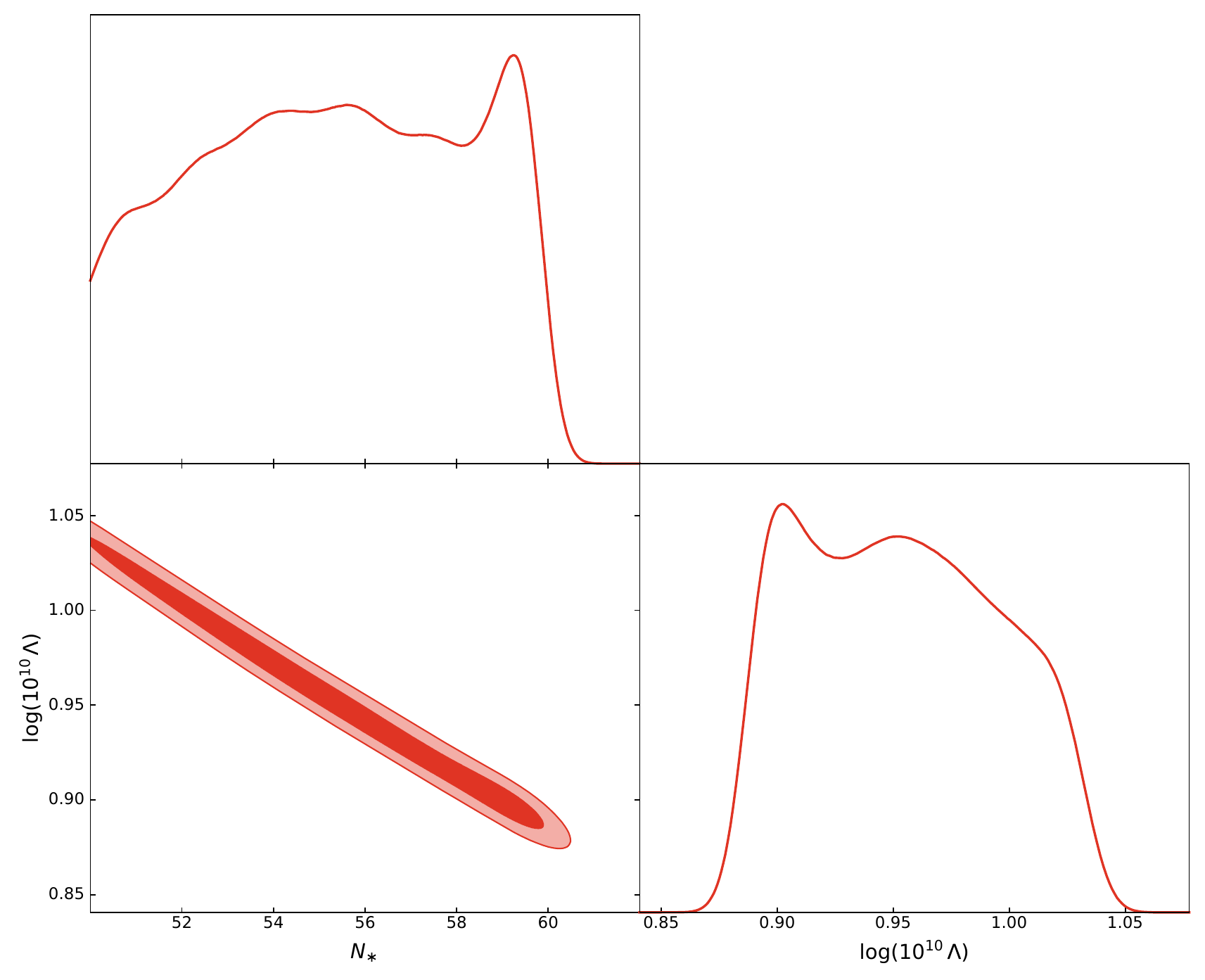}
\includegraphics[width=8.50cm]{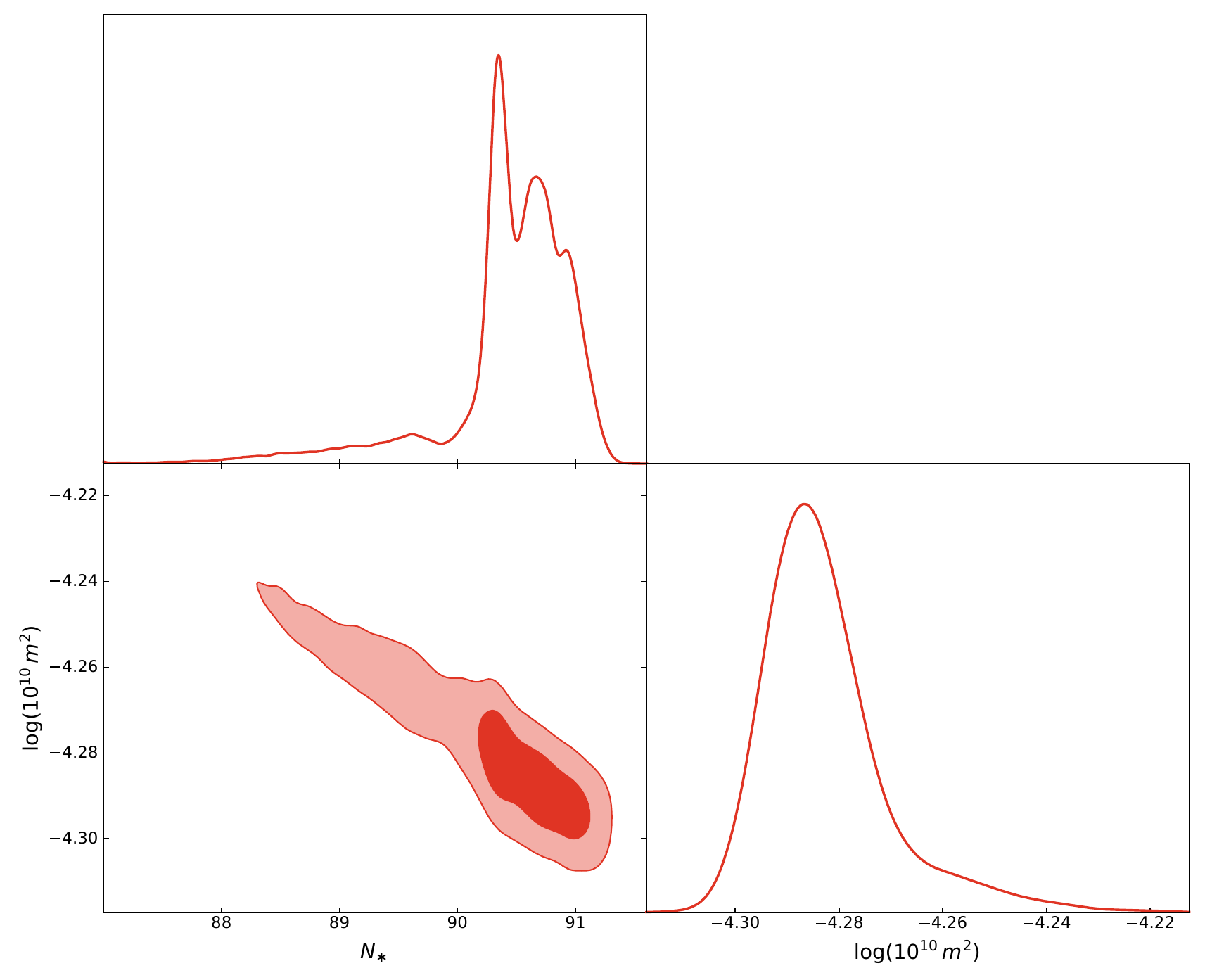}
\end{center}
\vskip -15pt
\caption{We have presented the contours of the marginalized posterior distribution 
(1-$\sigma$ regions in dark red and 2-$\sigma$ regions in light red) 
of the inflationary parameters $\Lambda$ or $m$ and $N_\ast$ in the models of SMIc 
(on the left) and PI (on the right).}\label{fig:contours1}
\end{figure*}

\begin{figure}
\includegraphics[width=15.00cm]{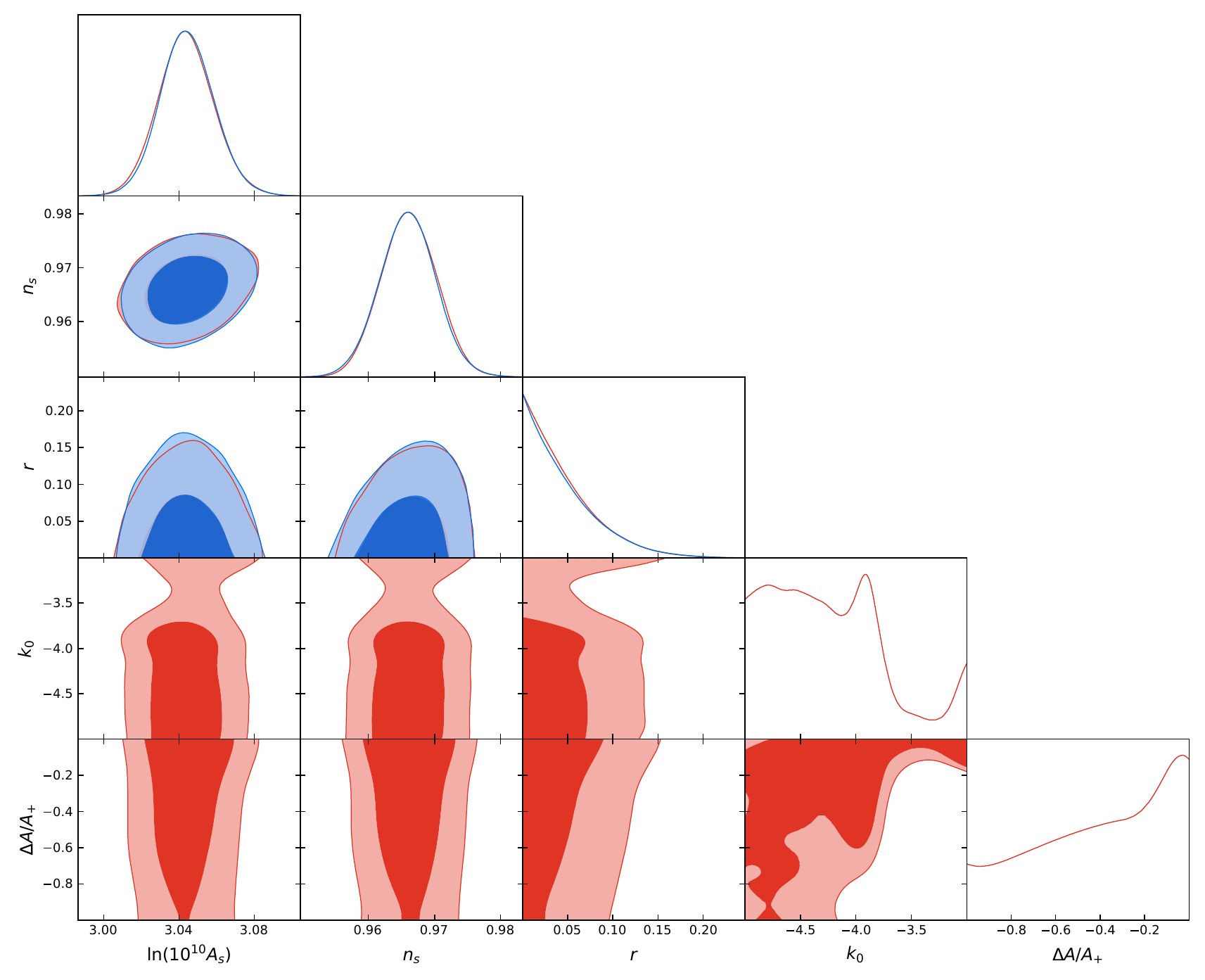}
\caption{We have illustrated the contours of the marginalized posterior
distribution of the parameters of the model SMII (in red) against the 
contours of the common parameters in the PL case (in blue), with a scale
invariant tensor amplitude assumed in both the cases.
Note that the contours on the parameters $\As$, $\ns$ and $r$ overlap 
considerably in the two cases.}\label{fig:contours2}
\end{figure}

\begin{figure}
\includegraphics[width=12.50cm]{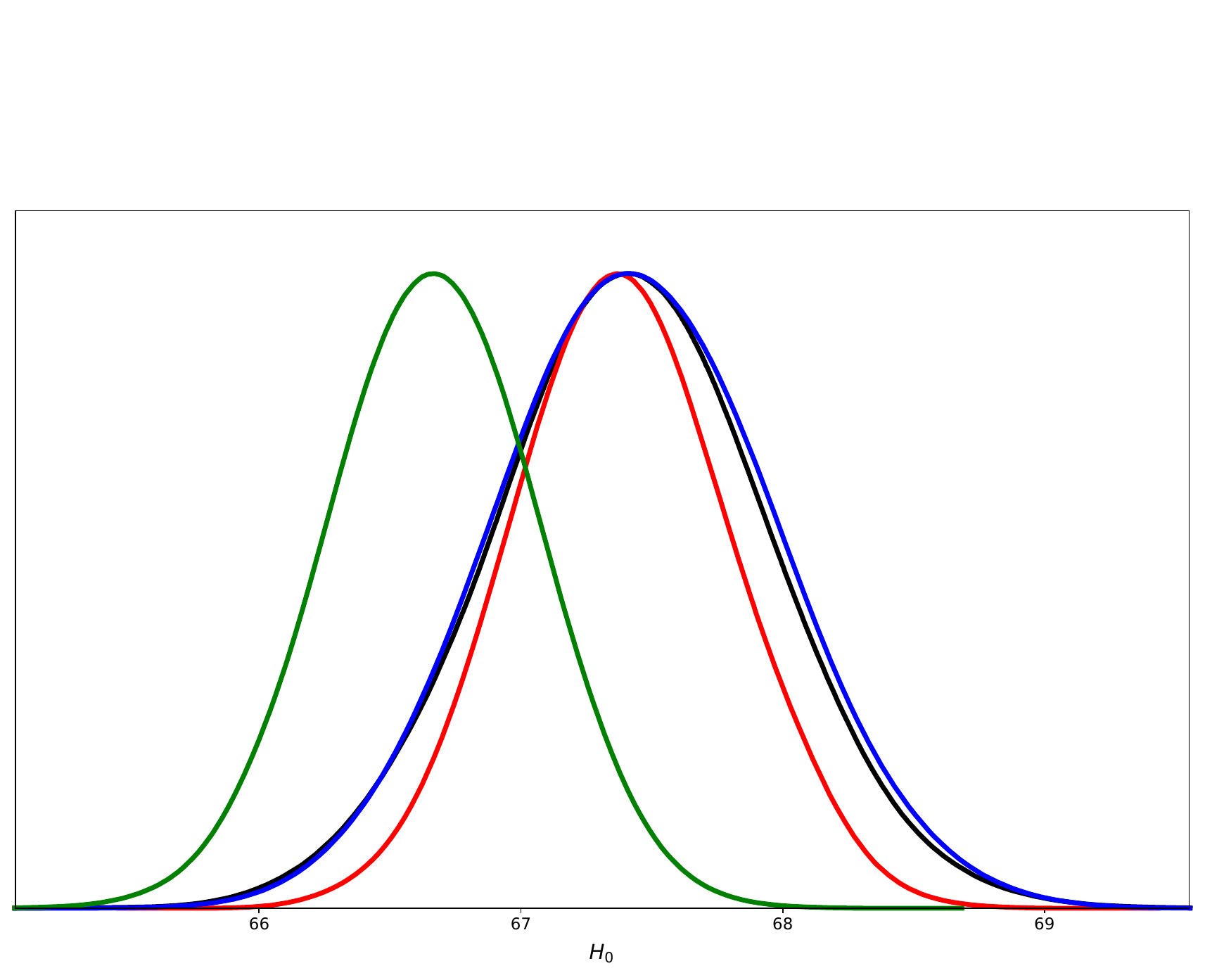}
\caption{We have presented the marginalized posterior distribution of 
the Hubble parameter $H_0$ arising in the PL case (in black) as well 
as in the inflationary models SMIc, SMII and PI (in red, blue and green).
It is clear that that the estimates of $H_0$ from SMIc and SMII match 
that of the PL case.
Whereas, the estimate from PI is slightly lower, which will actually 
aggravate the so-called Hubble tension.}\label{fig:H0_constraint}
\end{figure}

In this appendix, we shall present the confidence contours arrived at upon
comparing the inflationary models of our interest with the CMB data.
We shall focus on the three models SMIc, SMII and PI, which moderately 
improve the fit to the data [cf. Tab.~\ref{tab:chi2-bfv-recent}].
We find that the background cosmological parameters, viz. $\Omega_\mathrm{b}h^2$, 
$\Omega_\mathrm{c}\,h^2$, $\theta$ and~$\tau$ are constrained around the same 
region in the parameter space in cases of PL, SMIc and SMII. 
In case of PI, the contours differ slightly.
However, we find that the distributions are within 1-$\sigma$ intervals 
of each other.
In Fig.~\ref{fig:contours1}, we have illustrated the confidence contours 
for the inflationary parameters in the cases of SMIc and PI, models which
are described by the parameters $\Lambda$ and $m$, apart from $N_\ast$.
%%%%%%%%%%%%%%%%%%%%%%%%%%%%%%%%%%%%%%%%%%%%%%%%%%%%%%%%%%%%%%%%%%%%%%%%%%%%%%%%
%%%%%%%%%%%%%%%%%%%%%%%%%%%%%%%%%%%%%%%%%%%%%%%%%%%%%%%%%%%%%%%%%%%%%%%%%%%%%%%%
In Fig.~\ref{fig:contours2}, we have illustrated the constraints on all
the inflationary parameters in the model SMII. 
In the figure, we have also included the constraints on the inflationary
parameters in the PL case with a scale invariant tensor amplitude.
%%%%%%%%%%%%%%%%%%%%%%%%%%%%%%%%%%%%%%%%%%%%%%%%%%%%%%%%%%%%%%%%%%%%%%%%%%%%%%%%
%%%%%%%%%%%%%%%%%%%%%%%%%%%%%%%%%%%%%%%%%%%%%%%%%%%%%%%%%%%%%%%%%%%%%%%%%%%%%%%%
Lastly, in Fig.~\ref{fig:H0_constraint}, we have presented the marginalized 
posterior distributions on the Hubble parameter $H_0$, which is of considerable
interest today (in this context, see, for instance, the recent
reviews~\cite{DiValentino:2021izs,Freedman:2021ahq}).
%%%%%%%%%%%%%%%%%%%%%%%%%%%%%%%%%%%%%%%%%%%%%%%%%%%%%%%%%%%%%%%%%%%%%%%%%%%%%%%%
%%%%%%%%%%%%%%%%%%%%%%%%%%%%%%%%%%%%%%%%%%%%%%%%%%%%%%%%%%%%%%%%%%%%%%%%%%%%%%%%
We find that, while the mean values of $H_0$ in the cases of SMII and SMIc
roughly match the value in the standard PL case, the mean value is lower in
the case of PI which will aggravate the $H_0$ tension.
%%%%%%%%%%%%%%%%%%%%%%%%%%%%%%%%%%%%%%%%%%%%%%%%%%%%%%%%%%%%%%%%%%%%%%%%%%%%%%%%
\bibliographystyle{apsrev4-2}
\bibliography{kdi-v2}
%%%%%%%%%%%%%%%%%%%%%%%%%%%%%%%%%%%%%%%%%%%%%%%%%%%%%%%%%%%%%%%%%%%%%%%%%%%%%%%%
\end{document}